\documentclass[12pt]{article}
\usepackage[margin=1.25in]{geometry}
\usepackage{amsmath,amssymb,amsthm}
\usepackage[authoryear]{natbib}
\usepackage{mathtools,float}
\allowdisplaybreaks[4]
\usepackage[table,dvipsnames]{xcolor} 
\usepackage{booktabs}
\usepackage[T1]{fontenc}
\usepackage[latin9]{inputenc}
\usepackage{graphicx}
\usepackage{rotating}
\usepackage{comment}
\usepackage{url}
\usepackage{titletoc}
\usepackage{setspace}

\usepackage{subcaption}
\usepackage{float}

\definecolor{DarkBlue}{rgb}{0,0,0.6}
\usepackage{hyperref}
\hypersetup{
     colorlinks   = true,
     citecolor    = Maroon,
     urlcolor= DarkBlue,
     linkcolor=DarkBlue
}

\def\spacingset#1{\renewcommand{\baselinestretch}%
	{#1}\small\normalsize} \spacingset{1}

\begin{document}
\title{Green governments\thanks{First version: December 3, 2020. We are grateful to Toke Aidt, Zareh Asatryan, Stefan Bauernschuster, Christian Bj{\o}rnskov, Frank Bohn, Albrecht Bohne, Judson Boomhower, Gordon Dahl, Luisa D\"orr, Axel Dreher, Ottmar Edenhofer, Oliver Falck, Clemens Fuest, Sebastian Garmann, Kai Gehring, Robert Germeshausen, Klaus Gr\"undler, Jerg Gutmann, Carsten Hefeker, Friedrich Heinemann, Matthias Kalkuhl, Bj\"orn Kauder, Nicolas Koch, Manuela Krause, Tommy Krieger, Jana Lippelt, Markus Ludwig, Mathias Mier, Andreas Peichl, Karen Pittel, Rick von der Ploeg, Felix R\"osel, Fabian Ruthardt, Jeff Shrader, Guido Schwerdt, Edson Severini, Ulrich Wagner, Timo Wollmersh\"auser, Larissa Zierow, Katharina Zigova and the participants of the 2020 Silvaplana Political Economy Workshop, the CESifo Public Economics Conference 2020, the European Public Choice Society 2021, the Mannheim Conference on
Energy and the Environment 2021, and seminars at the ZEW Mannheim, the MCC Berlin and the ifo Institute Munich for comments. We are also grateful for comments when presenting our study at the Ministry of the Environment, Climate Protection and the Energy Sector in Baden-Wuerttemberg. 
Raphael de Britto Schiller, Tom Forntheil, Lea Fricke, Armin Hackenberger, Theresa Hailer, Niko Muffler, Maximilian Thomas, Tobias Urbin, and Timo Wochner provided excellent research assistance.}} 
\author{Niklas Potrafke\thanks{Department of Economics, University of Munich and Ifo Institute, Ifo Center for Public Finance and Political Economy, Poschingerstr.\ 5, D-81679 Munich. \url{potrafke@ifo.de}} \qquad Kaspar W\"uthrich\thanks{Department of Economics, University of California, San Diego, 9500 Gilman Dr.\ La Jolla, CA 92093; CESifo; Ifo Institute. \url{kwuthrich@ucsd.edu}}}

\date{\today}

\maketitle

\spacingset{1} 


\begin{abstract} 

We examine how Green governments influence environmental, macroeconomic, and education outcomes. We exploit that the Fukushima nuclear disaster in Japan gave rise to an unanticipated change in government in the German state Baden-Wuerttemberg in 2011. Using the synthetic control method, we find no evidence that the Green government influenced CO2 emissions or increased renewable energy usage overall. The share of wind power usage even decreased. Intra-ecological conflicts prevented the Green government from implementing drastic changes in environmental policies. The results do not suggest that the Green government influenced macroeconomic outcomes. Inclusive education policies caused comprehensive schools to become larger.

\medskip



\bigskip

\end{abstract}

\newpage

\onehalfspacing

\section{Introduction}
\label{sec: introduction}

Climate change is one of the ultimate challenges \citep[e.g.,][]{nordhaus2019}. In democracies, a major question is which political decision-makers are likely to successfully handle climate change. Green parties are expected and promise to foster renewable energies and improve environmental outcomes. Green parties have enjoyed tremendous electoral success in many countries (e.g., Austria, Finland, Germany, Iceland, Latvia, and New Zealand) over the last decades. In the wake of this success, they have gained more and more executive power and become part of state and national governments. 
However, little is known about what Green parties do when they are in office and lead governments. Which policies do they implement? How do they influence economic outcomes? Do they deliver on their promise to foster renewable energies and improve environmental outcomes? 

We provide the first causal evidence on how Green governments influence environmental, macroeconomic, and education outcomes. 
Our empirical strategy exploits that the Fukushima nuclear disaster in Japan gave rise to an unanticipated change in government in the German state Baden-Wuerttemberg (BW). On March 11, 2011, a tsunami following an earthquake destroyed the Fukushima Daiichi Nuclear Power Plant in Okuma (Japan), Fukushima Prefecture. Polling data show that the Fukushima nuclear disaster influenced the outcome of the state elections in BW on March 27, 2011. For the first time in history, a Green politician became the prime minister of a German state: Winfried Kretschmann became prime minister of the traditionally conservative state BW that had been governed by prime ministers from the conservative Christian Democratic Union (CDU) for 58 years until 2011. The Green party formed a Green-led coalition with the Social Democratic Party (SPD), which we refer to as \emph{Green government} for short. Importantly for our empirical strategy, the Fukushima nuclear disaster hardly changed election outcomes and coalition formation in other German states in 2011, and there has been no other Green prime minister. 

The unanticipated change of government provides an ideal setting for estimating the causal effect of a Green government on environmental, macroeconomic, and education outcomes. The Fukushima nuclear disaster opened a window of opportunity for changes especially in environmental and energy outcomes. We use the synthetic control (SC) method \citep{abadie2003economic,abadie2010synthetic,abadie2020jel} to construct a weighted average of other German states (referred to as ``synthetic'' BW), which measures how outcomes would have evolved in BW in the absence of a Green government. The SC method is well-suited for our purposes. It can be viewed as a generalization of the classical difference-in-differences method and provides a transparent data-driven approach for selecting control units \citep[see Section 4 of][for a discussion of the advantages of SC]{abadie2020jel}.

A key feature of our empirical design is that the Fukushima nuclear disaster only led to a Green government (our ``treatment'') in BW such that the treatment is unique. At the same time, the Fukushima nuclear disaster very likely influenced policies and outcomes in the other German states, making the synthetic BW ``greener''. Our identification strategy allows for such a direct effect.\footnote{Note that we are not using the Fukushima nuclear disaster as an instrumental variable for having a Green government (our treatment).} We only require this effect to be the same for synthetic BW and BW without the Green government.\footnote{Theoretically, SC is often motivated based on factor models with common shocks like the Fukushima nuclear disaster \citep[e.g.,][]{abadie2010synthetic,ferman2019properties}.}

We investigate three types of outcomes. First, we examine environmental and energy outcomes: the traditional focus of Green parties. Improving environmental outcomes --- for example, by promoting renewable energies --- is what voters expect from Green governments and what the Green party in BW had promised \citep{liststurm2006, gruene2011}.\footnote{Survey data before the 2011 (2016) state election show that 69\% (79\%) of the respondents believed that the Greens are competent in implementing suitable environmental policies \citep{tagesschau2011, tagesschau2016}.}  Environmental and energy policies have been the frontline policy issue of the Greens. However, we find no evidence that the Green government influenced CO2 emissions or increased energy usage from renewable energies overall.

An intriguing result of our study is that the share of wind power usage decreased relative to the estimated counterfactual. An interesting question is why. \emph{Intra-ecological conflicts} (nature and animal protection versus climate protection) and realities in public office are likely to have prevented the Green government from implementing drastic changes in environmental policies. Wind turbines are an important case in point to portray how environmental protection and animal protection conflict. The Green government also needed to handle ``not in my backyard'' movements, especially since it encouraged direct democracy. 
Moreover, some political projects, such as expanding wind energy, benefit from support at the local level, and the Greens in BW did not enjoy broad political majorities in the counties and municipalities.  

Second, we examine two key macroeconomic outcomes: GDP and the unemployment rate. The expected effect of a Green government on macroeconomic outcomes is ambiguous. On the one hand, Green parties have traditionally belonged to the leftwing political camp, and classical partisan theories predict that leftwing governments increase short-run GDP growth and employment \citep{hibbs1977political,chappell1986party,alesina1987macroeconomic}. On the other hand, the partisan theories were developed for traditional party systems in the 1970s and 1980s, ignoring Green governments. Moreover, policy uncertainty was pronounced because the Green government in BW was the first Green state government in Germany. Our results do not suggest that the Green government in BW influenced GDP and the unemployment rate. 

Finally, we investigate education outcomes. German state governments enjoy quite some leeway in designing education policies, which have been a controversial policy issue for a long time. The Green government implemented education policies promoting a more integrative school system (instead of a school system in which students are sorted based on their perceived abilities). We find that these policies drastically increased the number of students in comprehensive schools (\textit{Gesamtschulen}), including community schools (\textit{Gemeinschaftschulen}) in which students with varying abilities attend the same school. 

We contribute to the literature by presenting the first causal evidence on how Green governments perform in office. Our study relates to the literature studying Green parties in government \citep[e.g.,][]{neumayer2003are,knill2010do,cheon2013how,garmann2014do, schulze2021, jahn2022, toller2022}. The previous studies use panel data and report correlations between variables such as seat shares in parliaments and cabinets (when the Green parties are junior coalition partners) and outcomes such as environmental protection and CO2 emissions. However, such correlations cannot be interpreted as causal effects due to anticipation effects, reversed causality, and the endogeneity of changes in government. \citet{folke2014} employs discontinuities around seat thresholds in Swedish municipal councils to examine how Green parties influence environmental policies as measured by survey-based environmental rankings. We make two contributions to this literature. First, we exploit a plausibly exogenous and unanticipated change of government to estimate causal effects. Second, while the previous studies focus on Green parties as junior coalition partners or Green parties in parliament, we study the case of a Green government.

We also contribute to the partisan theories, which describe how governments' party composition influences economic policies and outcomes \citep{hibbs1977political,chappell1986party,alesina1987macroeconomic}. Empirical studies include \citet{ferreira09do}, \citet{gerber2011when}, \citet{fredriksson2011}, \citet{beland2015governors},
\citet{aidtetal2018},
and \citet{lind2020rainy}; see \citet{schmidt1996when} and \citet{potrafke2017partisan,potrafke2018government} for surveys. 
None of the previous studies investigates how Green governments influence economic policies and outcomes.

More broadly, our study contributes to the nexus between environmental policies/outcomes and macroeconomic outcomes. Examples include: the extent to which the environmental quality develops in line with economic growth --- the environmental Kuznets curve \citep[e.g.,][]{grossmankrueger1995}, how climate change influences economic growth \citep[e.g.,][]{kotzetal2021}, strategies to reduce CO2 emissions such as emission trading systems, carbon taxes and regulations \citep[e.g.,][]{aidt1998, aidt2010green, shapirowalker2018, andersson2019carbon, borensteinetal2019, barrage2020}. Another important question is how, in turn, environmental policies influence economic growth and employment \citep[e.g.,][]{metcalfstock2020}. Because environmental and energy policies are the frontline policy issues of Green parties, the nexus between environmental policies/outcomes and macroeconomic outcomes is quite likely to be influenced by Green governments. Our study shows how Green governments influence environmental and macroeconomic outcomes.

\section{The Fukushima disaster and the 2011 election}
\label{sec:2011 election and fukushima}

The 2011 state election in BW was historic. The conservative CDU set the prime minister for 58 years. The CDU had absolute majorities and formed single-party governments from 1972 to 1992. Before 1972 and after 1992, the CDU formed coalition governments either with the social-democratic SPD or the market-oriented FDP. The CDU had been the predominant party in BW for decades and enjoyed very comfortable political majorities. State elections take place every five years in BW. The party vote shares in the previous state election on March 26, 2006, were: CDU 44.2\%, FDP 10.7\%, SPD 25.2\%, and Greens 11.7\%. The CDU and the FDP formed a rightwing coalition government in June 14, 2006. Even a year before the 2011 state election, which we examine, the CDU and FDP had a comfortable majority. Polls from February 18, 2010 were: CDU 43\%, FDP 11\%, SPD 20\% and Greens 17\% (infratest dimap). 

The conservative CDU lost the election in 2011 against the Greens and the social-democratic SPD. What was more, for the first time in history, a green politician --- Winfried Kretschmann --- became prime minister of a German state. Several events influenced the outcome of the 2011 election but, importantly for our research design, the Fukushima accident tipped the scales.

The incumbent CDU/FDP-government lost support in the polls in the year 2010 for two main reasons. First, the then CDU/FDP-governments initiated the reconstruction of the main station in the state's capital Stuttgart (Stuttgart 21). The electorate was quite divided regarding the reconstruction of the main station. Rightwing voters were in favor of converting the terminal into an underground through station. Plans on how to reconstruct the main station had been discussed since 1985 \citep{wagschal2013was}. The state parliament approved the plan for Stuttgart 21 in 2006, a referendum against the plans was denied in 2007, and constructions started in 2010.
Citizens protested against the constructions. Violence escalated at the protests on September 30, 2010. There were conciliations in fall 2010, including temporary building freezes. Stuttgart 21 influenced the 2011 state election --- the CDU lost popularity and the Greens benefited. In November 2011, a referendum on whether the state government should withdraw from the project did, however, not receive a majority.

Second, the CDU replaced the prime minister in February 2010. Stefan Mappus succeeded G\"unther Oettinger. Stefan Mappus was an unpopular incumbent \citep[e.g.,][]{wehner2013historische}. His platforms on energy policy helped to make the Fukushima nuclear disaster a game-changer. In September 2010, the federal government decided to extend the run-time of Germany's nuclear power plants. The decision of the conservative and market-oriented federal government was made against the votes of the leftwing opposition in the German parliament. A prominent proponent was Stefan Mappus, who was in favor of nuclear energy and advocated the ``out of the nuclear phase-out''. However, after the Fukushima nuclear disaster, the federal government changed its position on nuclear energy and started promoting nuclear phase-outs. Stefan Mappus eventually gave in, losing credibility.

Due to the developments in the year 2010, the Greens and the SPD had a majority in the polls in January 2011. In February 2011, the popularity of the Greens started to decline. Support for the reconstruction of the main station in Stuttgart increased.\footnote{In February 2011, the reconstruction was supported by 43\% of the respondents of a representative survey (40\% in November 2010) and not supported by 35\% (39\% in November 2010) \citep{swr2011patt}.}
The conciliations in the late fall and the arbitration award by Heiner Geissler (a retired CDU politician) in December 2010 influenced parties' popularities. The CDU managed to portray the Greens as a destructive party that is opposing everything \citep{zeit2011patt}.\footnote{Representative surveys showed, for example, that Stuttgart 21 was perceived as being less important than education policies \citep{swr2011patt}. The CDU was also perceived as being by far the most competent party regarding economic policies. In a representative survey, citizens were asked which party is the most competent one regarding economic policies. The results were: CDU 42\%, SPD 11\%, Greens 6\%, FDP 3\%, no party 20\%, do not know 18\% \citep{forschungsgruppe2011politbarometer}.}
The overwhelming success of the Greens in the polls and the surprisingly poor performance of the CDU in the fall/winter of 2010 seemed to be exaggerated. When it got serious before the election in 2011, the traditionally conservative electorate in BW returned to support the conservative incumbent government.

\begin{figure}
    \caption{Election polls}
    \begin{center}
    \includegraphics[width=0.6\textwidth,trim=0 1cm 0 1cm]{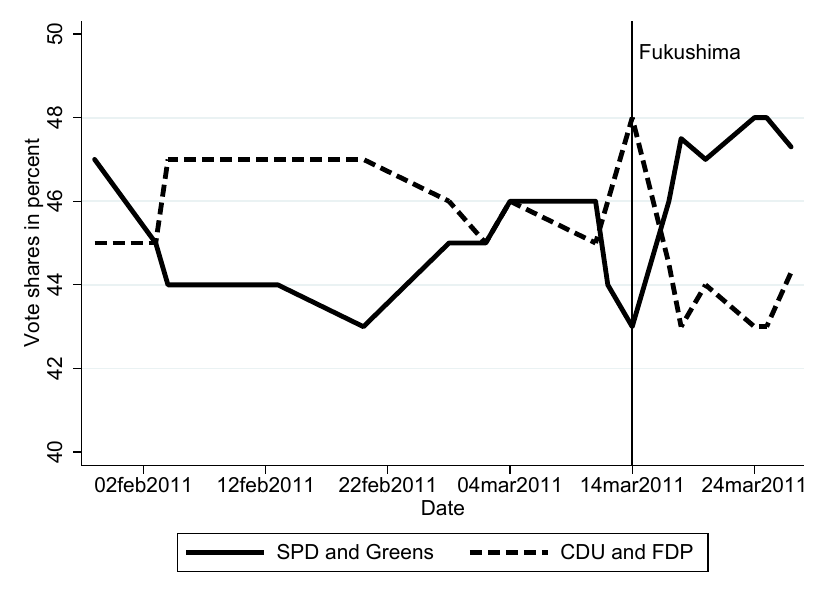}
    \end{center}
    \footnotesize{\textit{Notes:} The figure shows the polls of the four major pollsters infratest dimap, emnid, Forsa, Forschungsgruppe Wahlen in chronological order. We show all polls of the four major pollsters in chronological order to avoid any selection issues because the pollsters may have some political alignments. The latest poll that was started before the Fukushima disaster predicted: CDU 42\%, FDP 6\%, SPD 22\%, Greens 21\% (infratest dimap, published on March 14, polled March 10--12). The rightwing camp was in front by  48\% to 43\% (dashed line). The Fukushima nuclear disaster occurred on March 11. Things changed in the next poll that was also conducted by infratest dimap and published on March 17 (polled March 14--17): CDU 39\%, FDP 5.5\%, SPD 22\%, Greens 24\%.}
    \label{fig: vote shares}
\end{figure}

The predicted vote share of the Greens decreased to 19\% at the beginning of March 2011; the polls were published on March 2 and based on surveys conducted during February 21--25. The latest poll that was started before the Fukushima disaster predicted: CDU 42\%, FDP 6\%, SPD 22\%, Greens 21\% (infratest dimap, published on March 14, polled March 10--12). The rightwing camp was in front by 48\% to 43\% (Figure \ref{fig: vote shares}). The Fukushima nuclear disaster in Japan occurred on March 11. Things changed in the next poll, published on March 17 (polled March 14--17 by infratest dimap): CDU 39\%, FDP 5.5\%, SPD 22\%, Greens 24\%. The vote share of the CDU decreased by three percentage points, the vote share of the Greens increased by three percentage points. The two latest polls before the elections by Forsa and emnid reported that the vote shares of the leftwing camp were five percentage points higher than the vote shares of the rightwing camp (48\% to 43\%). The CDU received 38\%, the Greens 24\% and 25\%. The predicted vote shares of the SPD hardly changed. 
The state election took place on March 27. The change of power was sealed. Vote shares were: CDU 39\%, FDP 5.3\%, SPD 23.1\%, Greens 24.2\%. 

The Green party formed a coalition with the SPD. The coalition agreement was entitled ``The change begins''.\footnote{On party platforms, see Appendix \ref{app: party competition}.} Winfried Kretschmann became prime minister on May 12, 2011. The cabinet included the prime minister Kretschmann and twelve ministers: seven SPD ministers and five Green ministers (see Table \ref{tab: cabinet} in Appendix \ref{app:cabinet} for details). The SPD had more ministers than the Greens because Kretschmann became prime minister. In any event, two Green secretaries of state were also members of the government and were entitled to vote in the government. The Green party therefore had a majority in the government (eight Green against seven SPD politicians). They also had one more seat than the SPD in the state parliament.
Importantly for the Green party, Franz Untersteller was heading the ministry for the Environment, Climate Protection and Energy.

International media reported on how Fukushima tipped the scales. For example, the New York Times and the British Guardian headlined ``Merkel loses key German state on nuclear fears'' \citep{nyt2011merkel} and ``German Greens hail state victory in vote overshadowed by Fukushima'' \citep{guardian2011german}.
The Australian Sydney Morning Herald headlined ``German Greens on Fukushima high'' and explained ``A Green-led alliance with the Social Democrats won a four-seat majority in the state parliament of the southern state of Baden-Wurttemberg in a direct response to the Fukushima nuclear crisis'' \citep{smh2011german}.

\section{Empirical strategy}

\subsection{The synthetic control method}
\label{sec: synthetic control}
We use the SC method \citep{abadie2003economic,abadie2010synthetic,abadie2015comparative,abadie2020jel} to estimate the causal effect of the Green government on macroeconomic, education, and environmental outcomes in BW.\footnote{We contribute to the fast-growing literature that uses SC to estimate causal effects. Examples include: \citet{kleven2013taxation}, \citet{bohn2014did}, \citet{pinotti2015economic}, \citet{acemoglu2016value}, \citet{eliason2018can}, \citet{cunningham2018decriminalizing}, \citet{andersson2019carbon}, \citet{peri2019labor}, and \citet{potrafkeetal2020}.} The empirical analyses were performed in \texttt{Stata} \citep{stata2019} and \texttt{R} \citep{R2020}.

Let $j$ index German states and $t$ index time periods. The value $j=1$ corresponds to BW and $j=2,\dots,J+1$ index $J$ other German states that serve as controls. Let $Y_{jt}$ be the observed outcome of state $j$ in period $t$; see Section \ref{sec: results} for a description of the outcomes of interest. We adopt the potential outcomes framework \citep{neyman1923application,rubin1974estimating}. Let $Y_{jt}(0)$ and  $Y_{jt}(1)$ denote the potential outcomes without and with the treatment. In our context, the ``treatment'' is having a Green government.

BW is untreated for $t\le T_0$ and treated for $t>T_{0}$. The control states remain untreated for all periods. Thus, observed outcomes are related to potential outcomes as $Y_{jt}=D_{jt}Y_{jt}(1)+(1-D_{jt})Y_{jt}(0)$, where $D_{jt}=1\{j=1,t>T_0\}$. The new Green government took office on May 12, 2011. We therefore consider the year 2011 as the first treatment period such that $T_0=2010$. Doing so follows studies on partisan politics which assign a year in which a government changes to the government that was in power for at least six months \citep{potrafke2017partisan}. We examine data until 2015 because the next state election took place in March 2016. The Greens also won the 2016 elections and formed a coalition government with the conservative CDU.

We are interested in the causal effect of the change in government in BW after 2010:
\[
\alpha_t=Y_{1t}(1)-Y_{1t}(0),\quad t \in \{2011,\dots,2015\}
\]
Note that $Y_{1t}(1)$ (the potential outcome with a Green government) is observed in the post treatment period, whereas $Y_{1t}(0)$ (the potential outcome without a Green government) is fundamentally unobserved such that $\alpha_t=Y_{1t}(1)-Y_{1t}(0)=Y_{1t}-Y_{1t}(0)$.

To estimate $\alpha_t$, we need to estimate $Y_{1t}(0)$. We consider the following SC estimator:
\begin{equation}
\hat{Y}_{1t}(0)=\sum_{j=2}^{J+1}\hat{w}_{j}Y_{jt}(0)=\sum_{j=2}^{J+1}\hat{w}_{j}Y_{jt}, \label{eq:sc}
\end{equation}
where the second equality follows because $Y_{jt}(0)= Y_{jt}$ for $j\ge 2$ and all $t$ since the control states are untreated.
In equation \eqref{eq:sc}, we approximate the potential outcome of BW using a weighted combination of the contemporaneous (potential) outcomes of the other German states. We refer to this weighted combination as the ``synthetic BW''. 

We estimate the weights based on the pre-treatment data. Let $X_1,\dots,X_{J+1}$ denote vectors of predictors and define $X_0\equiv [X_{2},\dots,X_{J+1}]$. Different choices of predictors $X_j$ are possible. To mitigate concerns of specification searching, we use all pre-treatment outcomes and no additional covariates \citep[e.g.,][]{doudchenko2016balancing}.\footnote{We refer to \citet{botosaru2019role} and \citet{kaul2017synthetic} for a discussion of the role of additional covariates in SC settings.} The weights are obtained as 
\begin{eqnarray*}
\hat{w}\equiv(\hat{w}_2,\dots,\hat{w}_{J+1})&=&\arg\min_{w}\sqrt{(X_1-X_0w)'\Omega(X_1-X_0w)}\\
&\text{s.t.}& w\ge 0~ \text{and}~ \sum_{j=2}^{J+1}w_j=1.
\end{eqnarray*}
We implement SC using the Stata package \texttt{synth} \citep{synth}, which computes the matrix $\Omega$ using a data-driven regression-based method. We emphasize two important features of the SC weights \citep[][Section 4]{abadie2020jel}. First, due to the constraints imposed on the estimation problem, $\hat{w}$ will typically be a sparse vector (i.e., only contain few non-zero weights), which facilitates the interpretation of the synthetic BW. Second, the adding-up and positivity constraints preclude extrapolation beyond the support of the control data. 

To make inferences, we employ the widely-used permutation method of \citet{abadie2010synthetic}; see also \citet{firpo18synthetic} and \citet[][Section 3.5]{abadie2020jel} for further discussions.\footnote{Our post-treatment period only comprises five years such that inference methods relying on many post-treatment periods such as \citet{chernozhukov2019ttest} and \citet{li2019statistical} are not suitable here.} In Section \ref{sec: robustness}, we show that our results are robust to using the recently proposed conformal inference procedure of \citet{chernozhukov2019exact}.\footnote{To implement the conformal inference method, we use the \texttt{R}-package \texttt{scinference} (\url{https://github.com/kwuthrich/scinference}).}

\subsection{Choice of donor pool}
\label{subsec: choice of donor pool}

We use data going back to 1992, the period after the German unification. 
To construct the control group (referred to as ``donor pool''), we use other German states.\footnote{Germany has 16 states: ten West German states and six East German states (including Berlin).} BW had been governed by the conservative CDU for 58 years before treatment. Therefore, we choose as our donor pool all German states that had the conservative CDU in government, at least for some years during our pre-treatment period. Our final donor pool includes all other German states except Rhineland-Palatinate.\footnote{The CDU was governing in Rhineland-Palatinate until spring 1991, just before our pre-treatment period begins in 1992. The SPD won the state elections in spring 1991 and has been ruling in coalitions with the market-oriented FDP and the Greens since 1991.} There is one other German state, Bavaria, that had also been governed by the conservative CSU (the CDU's sister party) for decades before treatment. BW and Bavaria were quite similar along many relevant dimensions including economic performance and preferences of the electorate before treatment such that Bavaria is a natural control unit. 
Indeed, for some outcome variables of interest (e.g.,  wind energy and unemployment rate), Bavaria receives an SC weight of (almost) one such that our SC analysis is very similar to difference-in-differences using Bavaria as the control unit.

Our treatment is having a Green government. When Bavaria receives much weight of the synthetic BW (e.g., wind energy usage and unemployment rate), our estimates capture the difference between a Green and a Christian-conservative/market-oriented government. When other states also receive positive weight, our estimates are causal effects relative to a counterfactual with different types of non-Green governments.

It is important to ensure that none of the control units is treated. Our treatment is unique. There has been no other Green prime minister in Germany until now, except for Winfried Kretschmann in BW. The Greens have not had as much executive power in any German state as they have in BW.\footnote{The Greens have been a junior coalition partner since the 1980s in many German states, but they have never been leading a state government. They were also a junior coalition partner at the national level \citep[e.g.,][]{zohlnhoefer2004}.} The Fukushima nuclear disaster very likely influenced parties' platforms and governments' policies in the other states as well. Only in BW, however, the Fukushima nuclear disaster gave rise to a Green government. Importantly, our identification strategy allows for the Fukushima nuclear disaster to have a direct effect on the outcomes of interest. We only require this direct effect to be the same for BW and synthetic BW without a Green government.\footnote{Theoretically, SC is typically motivated based on factor models for the potential outcome absent the treatment, $Y_{jt}(0)$ \citep[e.g.,][]{abadie2010synthetic,ferman2019properties}. Such factor models allow for common shocks like the Fukushima nuclear disaster with different impacts on different control units, provided that the weighted average impact is the same as for the treated unit.}

\section{Results}
\label{sec: results}

\subsection{Energy and environmental outcomes}

Energy and environmental policies have been the frontline policy issue of the Greens. Voters expect Green governments to improve energy and environmental outcomes. German state governments influence energy and environmental outcomes, for example, by promoting and regulating energy technologies \citep{wurster2017energiewende}. Promoting includes subsidizing individual energy technologies such as thermal energy. State governments regulate energy technologies and advertise areas to build wind turbines. 
They also influence the construction of power plants as well as power and gas lines. However, constructions of power plants and power and gas lines are typically long-term projects, the effects of which are unlikely to materialize within the five-year post-treatment period that we consider. 

In BW, the previous CDU-led government landed a political coup in 2010 that helped to influence energy policies a great deal. The CDU-led government repurchased stocks (45.01\%) of the major energy provider in BW from the French energy provider Electricite de France. The Green government thus possessed opportunities to influence energy outcomes directly  \citep{wurster2017energiewende}. In particular, the Greens held the ministry for Environment, Climate Protection and Energy (cf.\ Table \ref{tab: cabinet}). 

We expect Green governments to use renewable energy carriers such as wind energy to decrease CO2 emissions. Indeed, the Green party had promised to do so in their 2011 election program \citep[][p.\ 31f.]{gruene2011}.
We consider six outcomes: CO2 emissions, mineral oil, coal, renewable energies, wind energy, and solar energy.\footnote{We also investigated water energy usage. However, the SC method provided a very poor pre-treatment fit for water energy usage.} The energy outcomes are measured as a share of primary energy usage. Primary energy usage covers the energy content of all carriers within a state (industry, traffic, energy consumption of households, etc.) including mineral oil, coal, nuclear energy, as well as renewable energies.\footnote{Table \ref{tab: depvars} lists all dependent variables, data sources, and the time period considered based on data availability.} 
We do not consider nuclear energy because the federal government decided to abolish nuclear energy after the Fukushima disaster (cf.\ Section \ref{sec:2011 election and fukushima}). The data are from the State Working Committee for Energy Balances --- the agency of the German states that compiles and provides the data of energy balances in the German states.

\begin{figure}[H]
    \caption{Spaghetti graphs: energy outcomes}
    \begin{center}
        \vspace{-0.4cm}
    \begin{subfigure}[b]{0.325\textwidth}
         \centering
         \caption{CO2 emissions}
         \includegraphics[width=\textwidth]{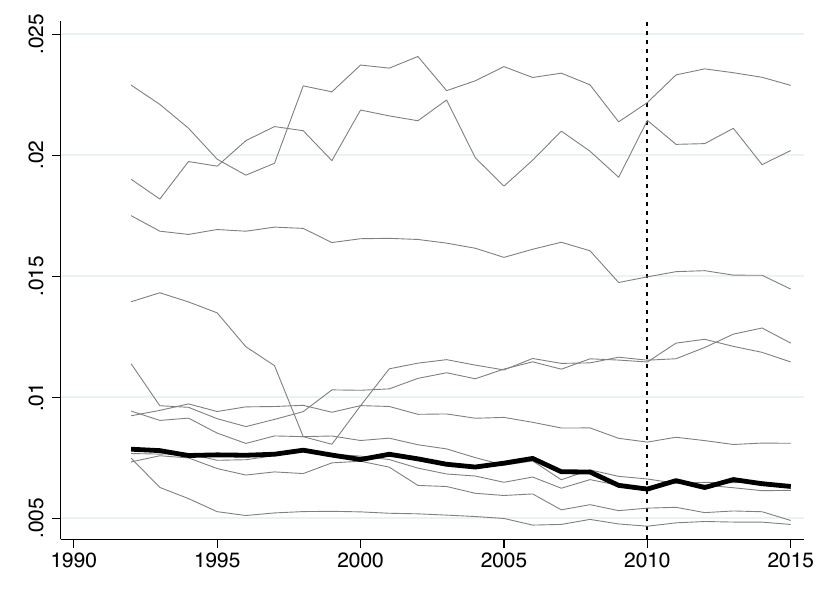}
    \end{subfigure}
    \hfill
    \begin{subfigure}[b]{0.325\textwidth}
         \centering
         \caption{Mineral oil}
         \includegraphics[width=\textwidth]{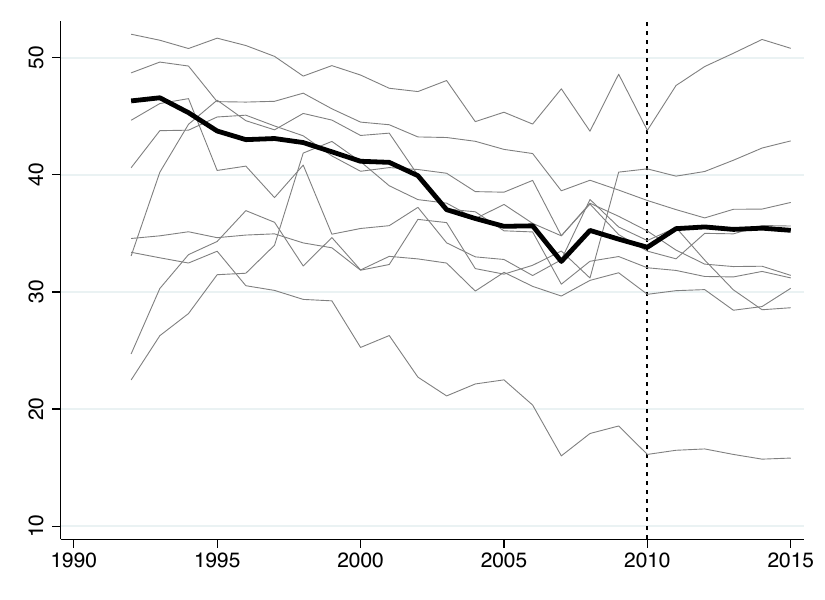}
         \end{subfigure}
    \begin{subfigure}[b]{0.325\textwidth}
        \centering
        \caption{Coal}
        \includegraphics[width=\textwidth]{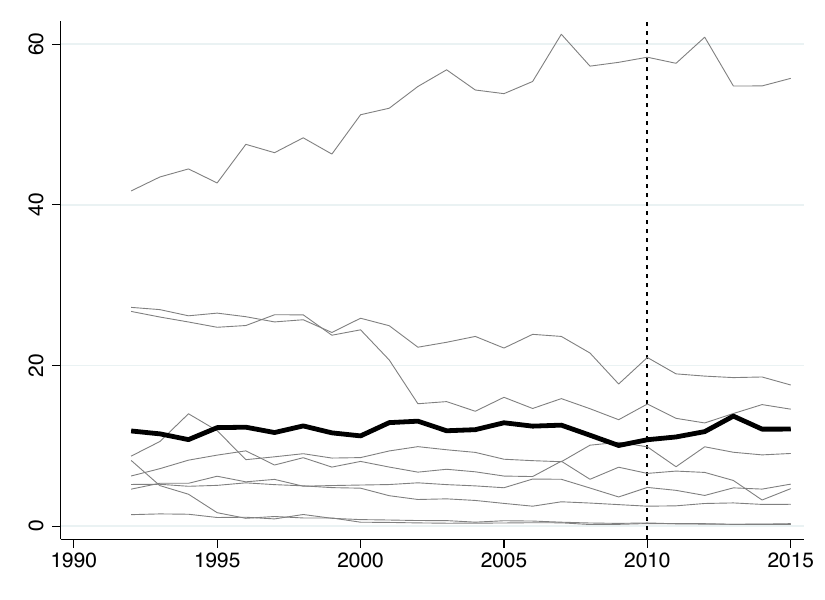}
    \end{subfigure}
    \begin{subfigure}[b]{0.325\textwidth}
        \centering
        \caption{Renewable energies}
        \includegraphics[width=\textwidth]{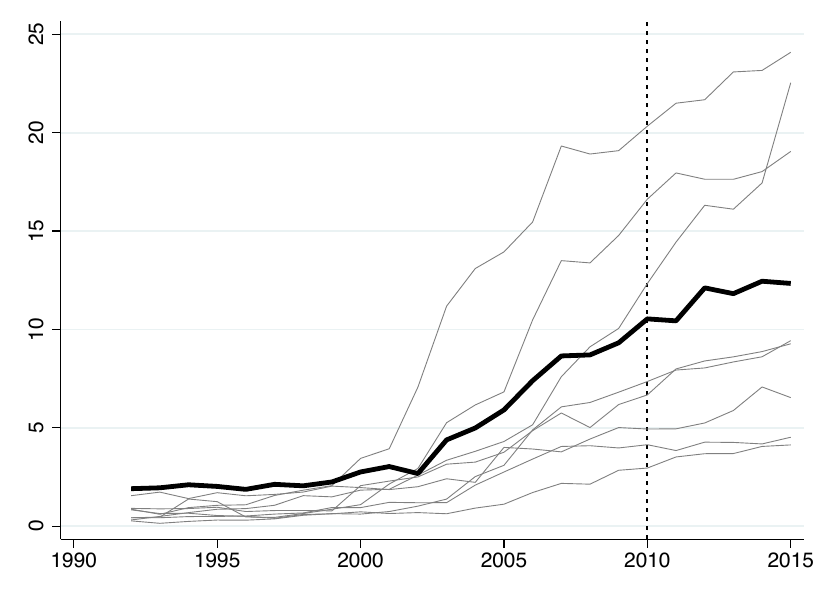}
    \end{subfigure} 
    \begin{subfigure}[b]{0.32\textwidth}
        \centering
        \caption{Wind energy}
        \includegraphics[width=\textwidth]{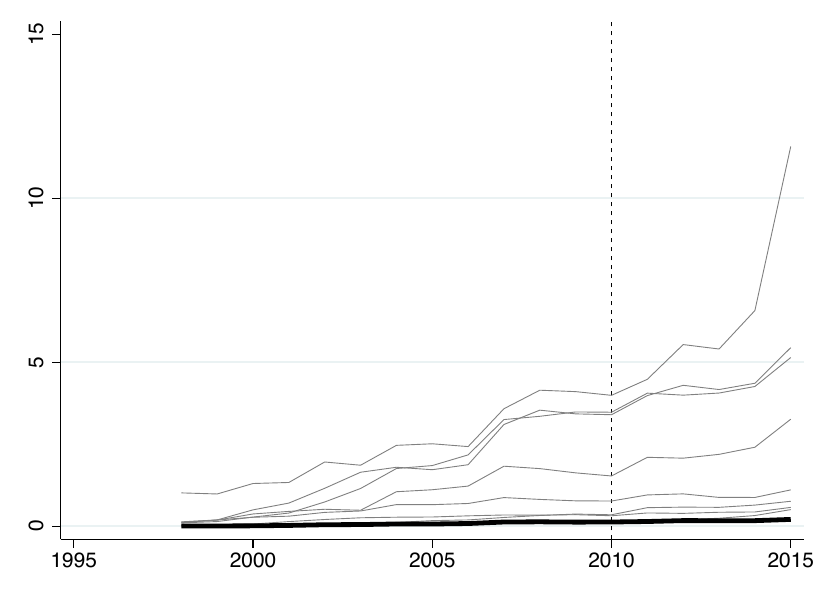}
    \end{subfigure}   
    \begin{subfigure}[b]{0.325\textwidth}
        \centering
        \caption{Solar energy}
        \includegraphics[width=\textwidth]{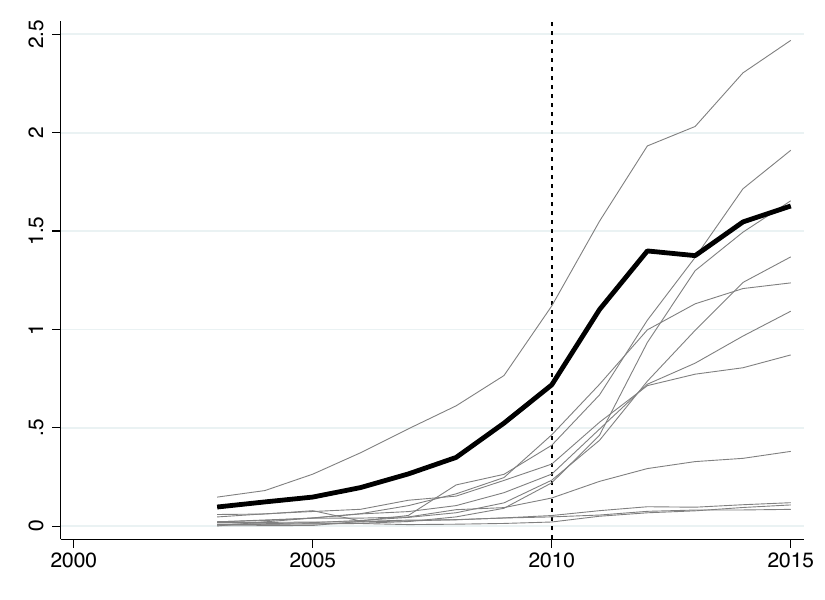}
    \end{subfigure} 
    \end{center}     
	\footnotesize{\textit{Notes:} CO2 emissions are measured in 1000t per inhabitant. The energy outcomes are measured as a share of primary energy usage. The data are from the State Working Committee for Energy Balances and the Federal Environment Agency.}
\label{fig:head_med_energy}
\end{figure}

Figure \ref{fig:head_med_energy} shows the raw data for BW (thick line) and the states in the donor pool. For all outcomes, there is substantial heterogeneity between the states. The SC method is particularly useful in this context. It allows for choosing suitable control states in a transparent and data-driven way. 

Table \ref{tab: tableenergyweights} shows the estimated SC weights for the individual outcomes. Bavaria is quite important: it gets, for example, 33\% (CO2 emissions), 52.7\% (mineral oil) and 94.8\% (wind energy) of the weight in the synthetic BW. This result is reasonable since Bavaria in many ways is a ``natural'' control state for BW. For example, it is the only state governed by the conservative CSU (the CDU's sister party that is only running in Bavaria) before and after treatment in BW, and population sizes are quite comparable (Bavaria has 13 and BW 11 million inhabitants). The results in Table \ref{tab: tableenergyweights} also demonstrate a key feature of SC: the SC method is adaptive and may well assign a very large weight to a single control state (i.e., essentially select a single control state) if it provides a good approximation to the counterfactual.

\begin{table}[H]
\tiny
\centering
\caption{SC weights: energy outcomes}

\begin{tabular}{lcccccc}
\\
\toprule
          & \multicolumn{1}{c}{\textbf{CO2 emissions}}
          & \multicolumn{1}{c}{\textbf{Mineral oil}}
          & \multicolumn{1}{c}{\textbf{Coal}} 
          & \multicolumn{1}{c}{\textbf{Renewable energies}} 
          & \multicolumn{1}{c}{\textbf{Wind energy}} 
          &  \multicolumn{1}{c}{\textbf{Solar energy}}

          \\ \cmidrule{2-7}     \\
    Brandenburg              & 0      & 0     & 0.593 & 0.229     & 0         & 0     \\
    Berlin                   & 0.095  & 0     & 0     & 0         &           & 0     \\
    Bavaria                  & 0.330  & 0.527 & 0     &           & 0.948     & 0.456 \\
    Bremen                   & 0      & 0.179 & 0.106 & 0.629     & 0         & 0     \\
    Hesse                    & 0.330  & 0.064 & 0     & 0         &           & 0.184 \\
    Hamburg                  &        &       &       &           &           & 0     \\
    Mecklenburg-Pommerania   &        &       &       &           &           &       \\
    Lower-Saxony             &        &       &       &           &           &       \\
    North Rhine-Westphalia   & 0      & 0     & 0.137 & 0         & 0.044     & 0     \\
    Schleswig-Holstein       & 0      & 0.123 & 0     & 0         & 0         & 0.087 \\
    Saarland                 &        &       &       &           &           &       \\
    Saxony                   & 0      & 0     & 0.163 & 0         & 0         & 0     \\
    Saxony-Anhalt            & 0      &       &       &           & 0.008     & 0     \\
    Thuringia                & 0.245  & 0.107 & 0     & 0.141     & 0         & 0.273 \\
\\
\bottomrule
\end{tabular}
\label{tab: tableenergyweights}
\end{table}

\begin{figure}[H]
	\caption{SC results: CO2, mineral oil, coal}
	
	\vspace{-0.4cm}
	\begin{center}
	\begin{subfigure}[b]{\textwidth}
	\caption{CO2 emissions in 1000 tons per inhabitant}
	    \begin{center}
		\includegraphics[width=0.325\textwidth,trim=0 0cm 0 1.5cm]{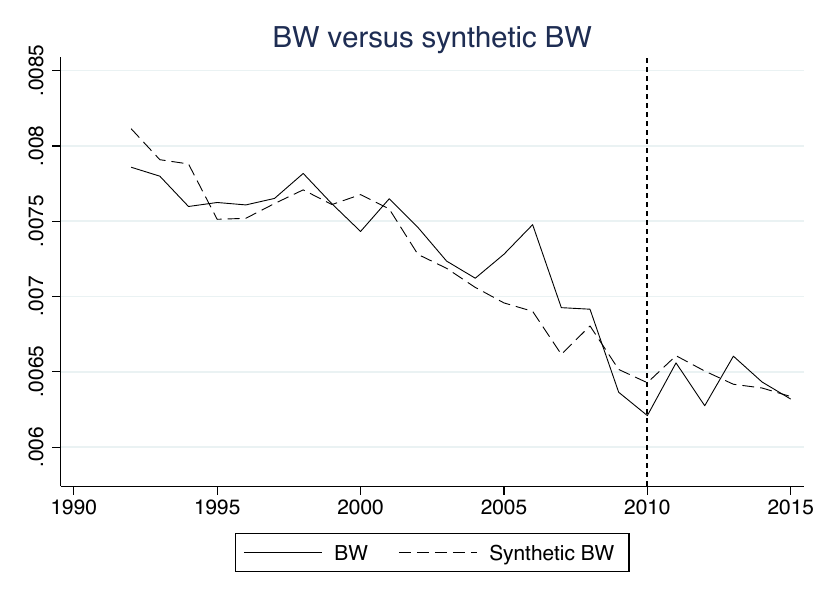}
		\includegraphics[width=0.325\textwidth,trim=0 0cm 0 1.5cm]{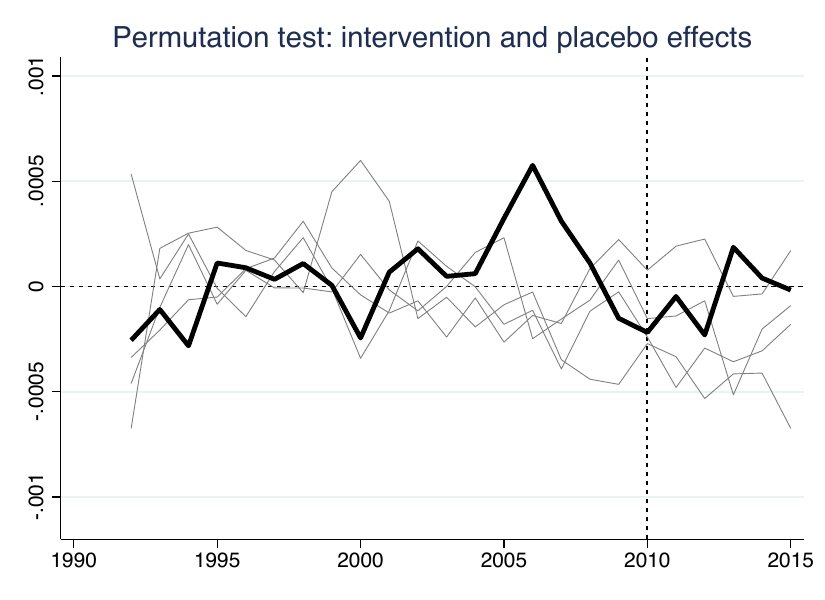}
		\includegraphics[width=0.325\textwidth,trim=0 0cm 0 1.5cm]{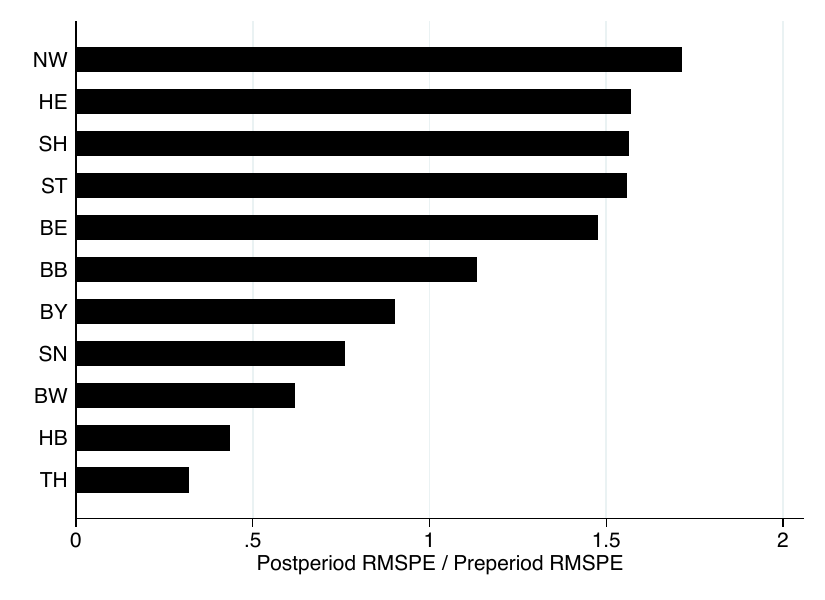}
	\end{center}
	\end{subfigure}
		\vspace{0.1cm}	
	\begin{subfigure}[b]{\textwidth}
	\caption{Mineral oil}
		\includegraphics[width=0.325\textwidth,trim=0 0cm 0 0cm]{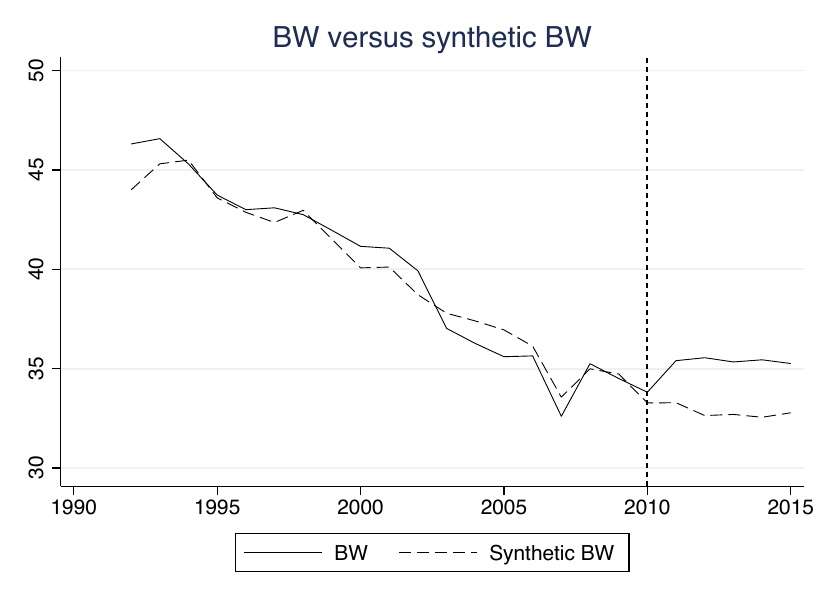}
		\includegraphics[width=0.325\textwidth,trim=0 0cm 0 0cm]{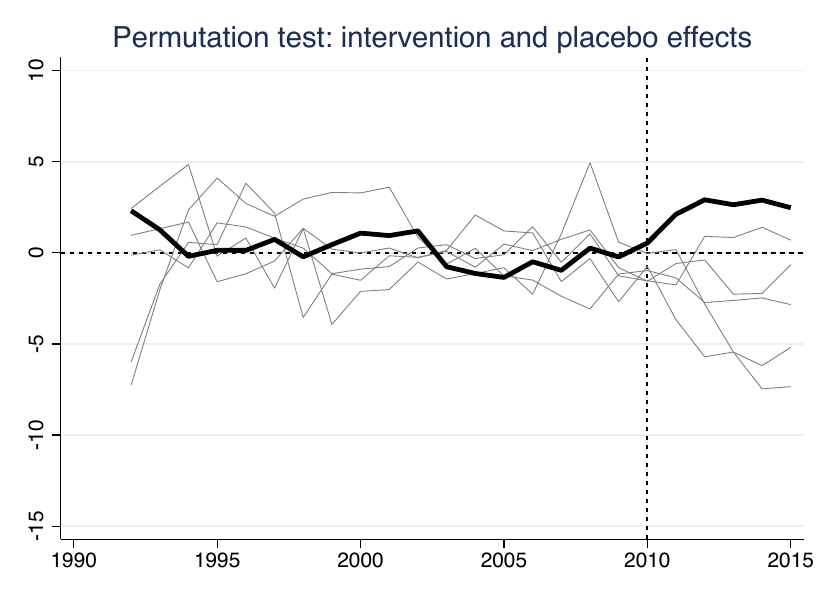}
		\includegraphics[width=0.325\textwidth,trim=0 0cm 0 0cm]{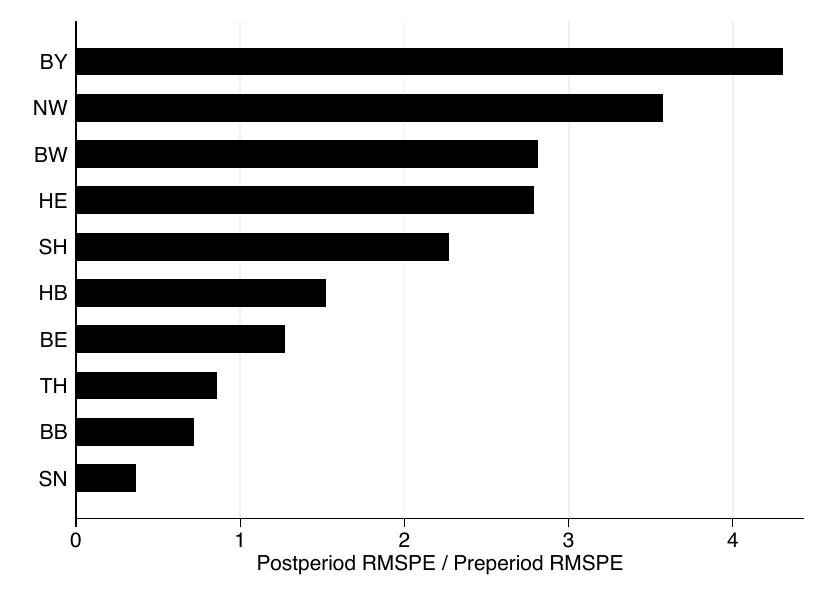}
	\end{subfigure}
		\vspace{0.1cm}
	\begin{subfigure}[b]{\textwidth}
	\caption{Coal}
		\includegraphics[width=0.325\textwidth,trim=0 0cm 0 0cm]{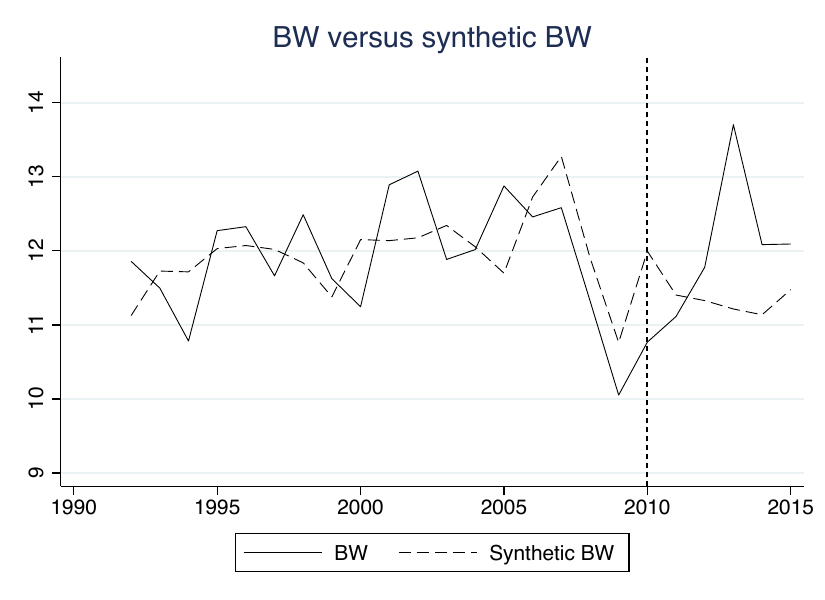}
		\includegraphics[width=0.325\textwidth,trim=0 0cm 0 0cm]{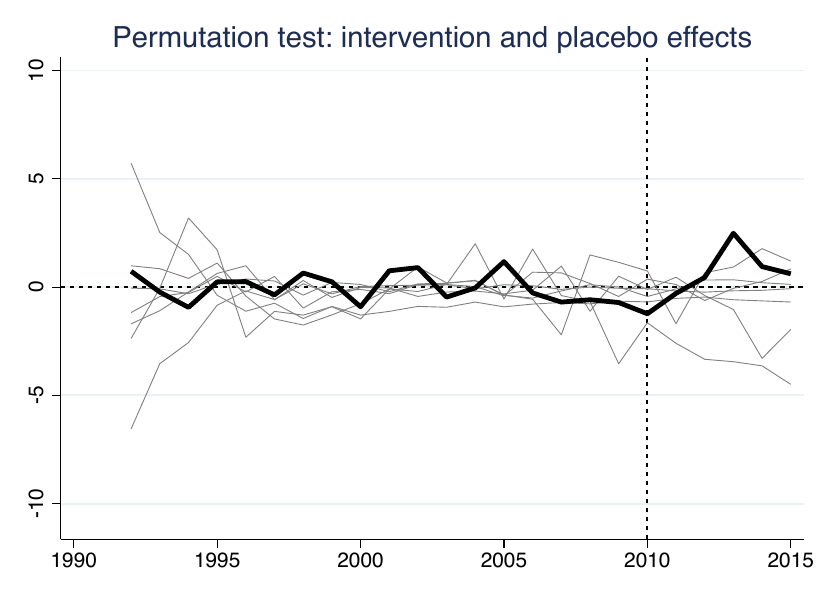}
		\includegraphics[width=0.325\textwidth,trim=0 0cm 0 0cm]{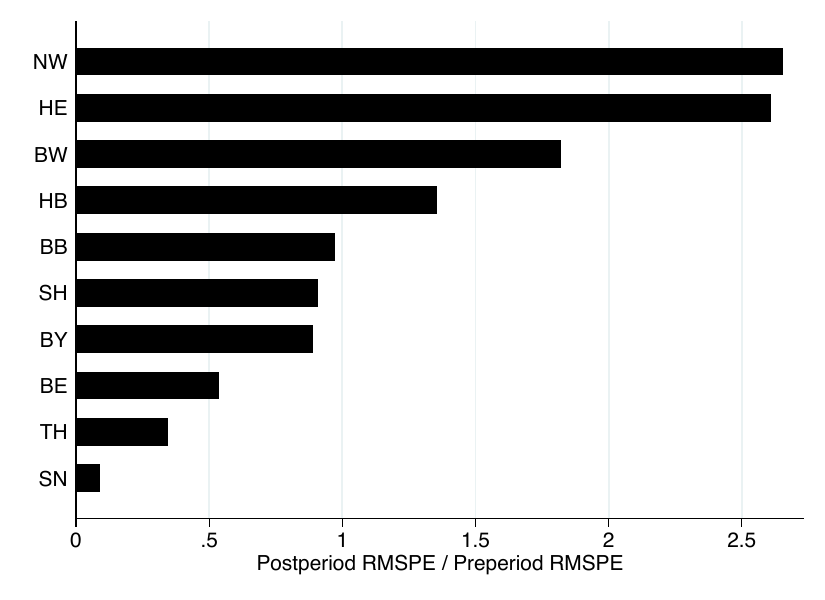}
	\end{subfigure}
	\end{center}
	\footnotesize{\textit{Notes:} The center figures exclude states for which the pre-treatment MSPE is at least 10 times larger than BW's pre-treatment MSPE. The data are from the State Working Committee for Energy Balances.}
	\label{fig: CO2 oil coal}
\end{figure}

Figure \ref{fig: CO2 oil coal} shows the SC results for CO2 emissions
(as measured in 1000 tons per inhabitant), the share of mineral oil, and the share of coal. How the Green government influenced the individual energy and environmental outcomes is measured by the gap between the individual energy and environmental outcomes in BW and synthetic BW after treatment. We find that, for example, mineral oil usage was somewhat higher in BW than in synthetic BW after treatment, whereas CO2 emissions were hardly lower in BW than in synthetic BW after treatment.

To assess the statistical significance of the results, we rely on the widely-used permutation inference approach of \citet{abadie2010synthetic}. We assign the treatment iteratively to every state in the donor pool. For some control states, the SC method does not deliver good pre-treatment fits. Therefore, we exclude states for which the pre-treatment MSPE (mean squared prediction error) is at least 10 times larger than BW's pre-treatment MSPE. The center subfigures of Figures \ref{fig: CO2 oil coal} present the results, allowing for a visual comparison of the gap in BW and the permutation distribution of placebo gaps. 

Clearly, the choice of the MSPE cutoff of 10 is arbitrary and one could, for example, also choose 5 or 20 as in \citet{abadie2010synthetic}. Therefore, we report the ratio of post-treatment root MSPE (RMSPE) to pre-treatment RMSPE, following \citet{abadie2015comparative}. A large ratio indicates a rejection of the null hypothesis that the Green government had no impact whatsoever. For CO2 emissions, the ratio for BW is the ninth largest among all states. The implied permutation $p$-value is $p = (J+1)^{-1}\sum_{j=1}^{J+1}1\left\{r_j\ge r_1\right\}=9/11$, where $r_j$ is the RMSPE ratio for unit $j$, and $j=1$ indexes BW. Thus, the results do not suggest that the Green government influenced CO2 emissions. We do not find a significant effect of the Green government on mineral oil usage either: the RMSPE ratio for BW is the third largest among all states. Coal usage gives rise to high CO2 emissions.\footnote{We also investigated lignite usage. For lignite usage, Bremen needs to be excluded from the donor pool due to industrial issues that would confound our analysis (manufacturing industries in Bremen started replacing coal with lignite in 2011). After excluding Bremen from the donor pool, the SC method did not provide a satisfactory pre-treatment fit.}
Therefore, the Green government is expected to decrease coal usage. However, Figure \ref{fig: CO2 oil coal} shows that the Green government even somewhat increased the share of coal, albeit the effect is not significant.

\begin{figure}[H]
	\caption{SC results: renewable energies}
	
	\vspace{-0.4cm}
	\begin{center}
	\begin{subfigure}[b]{\textwidth}
	     \caption{Renewable energies}
	    \begin{center}
		\includegraphics[width=0.325\textwidth,trim=0 0cm 0 1cm]{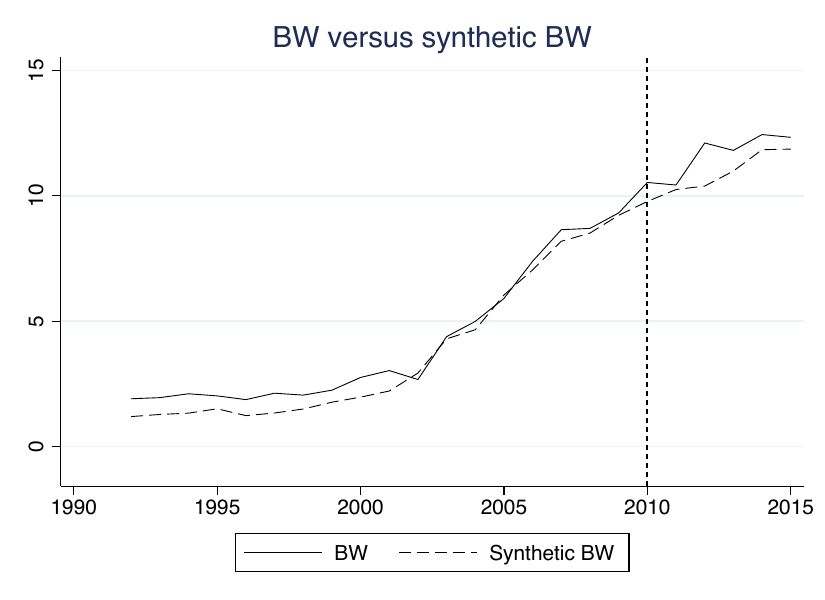}
		\includegraphics[width=0.325\textwidth,trim=0 0cm 0 1cm]{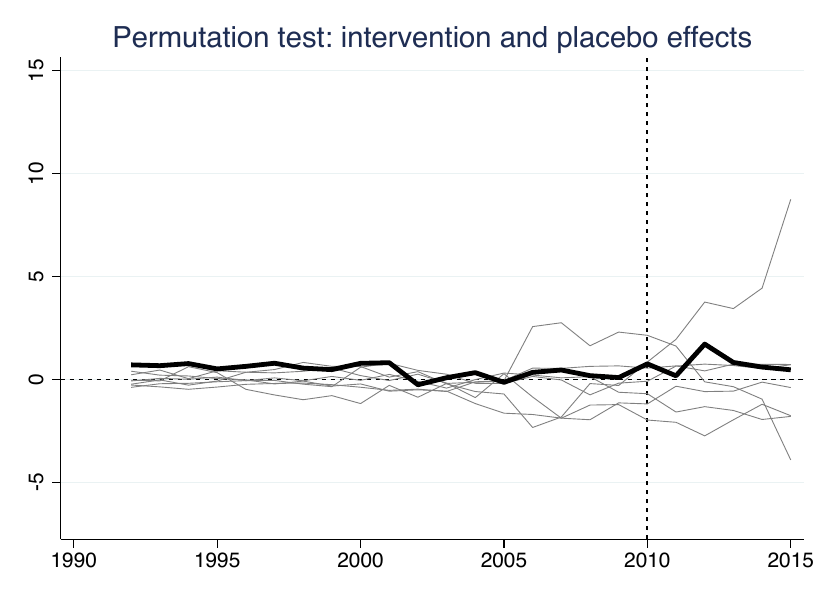}
		\includegraphics[width=0.325\textwidth,trim=0 0cm 0 1cm]{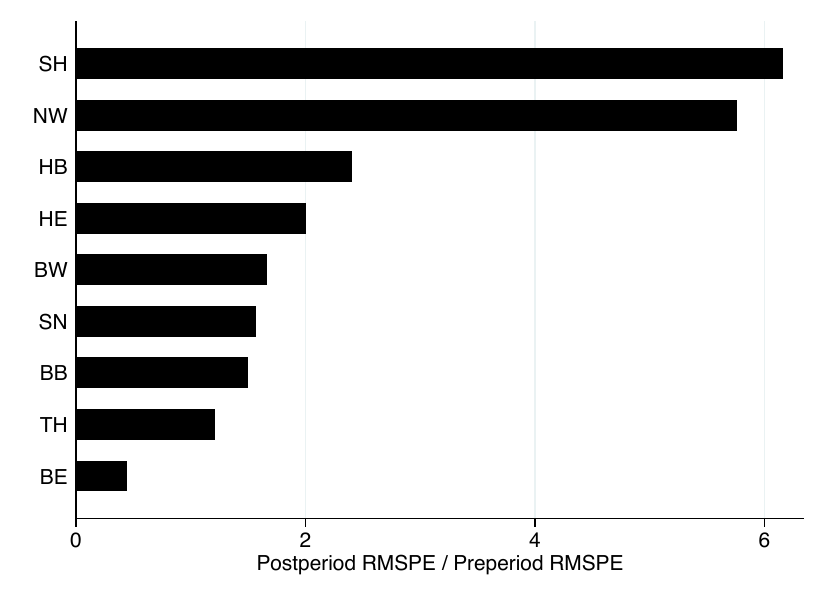}
	\end{center}
	\end{subfigure}
		\vspace{0.1cm}	
    \begin{subfigure}[b]{\textwidth}
	\caption{Solar energy}
	    \includegraphics[width=0.325\textwidth,trim=0 0cm 0 0cm]{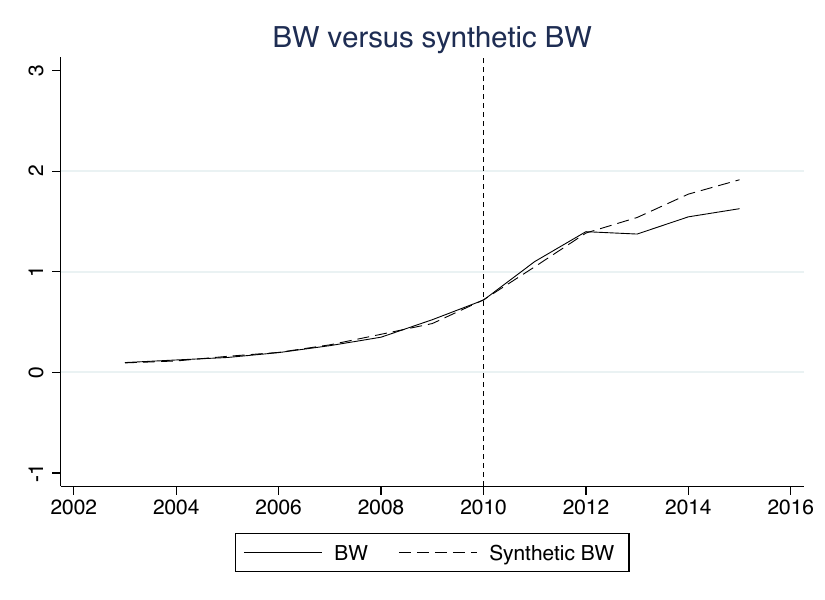}
		\includegraphics[width=0.325\textwidth,trim=0 0cm 0 0cm]{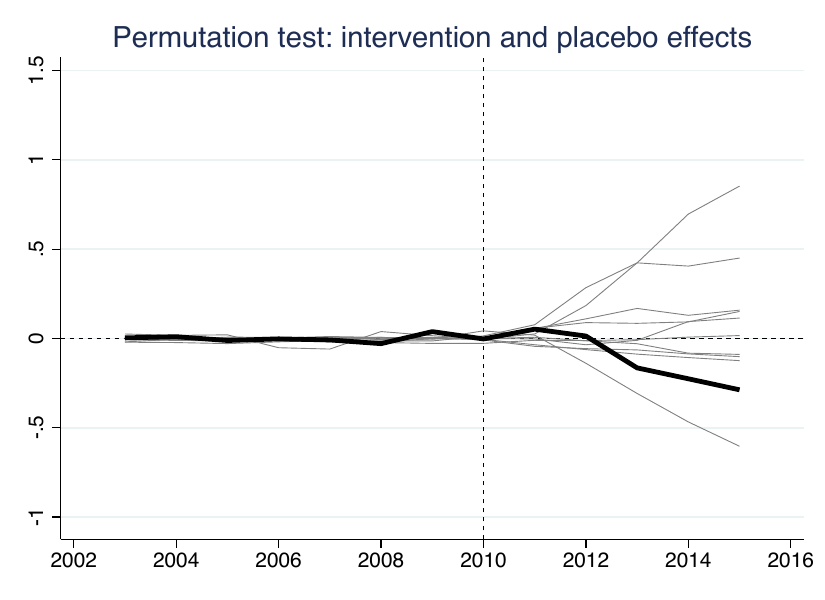}
		\includegraphics[width=0.325\textwidth,trim=0 0cm 0 0cm]{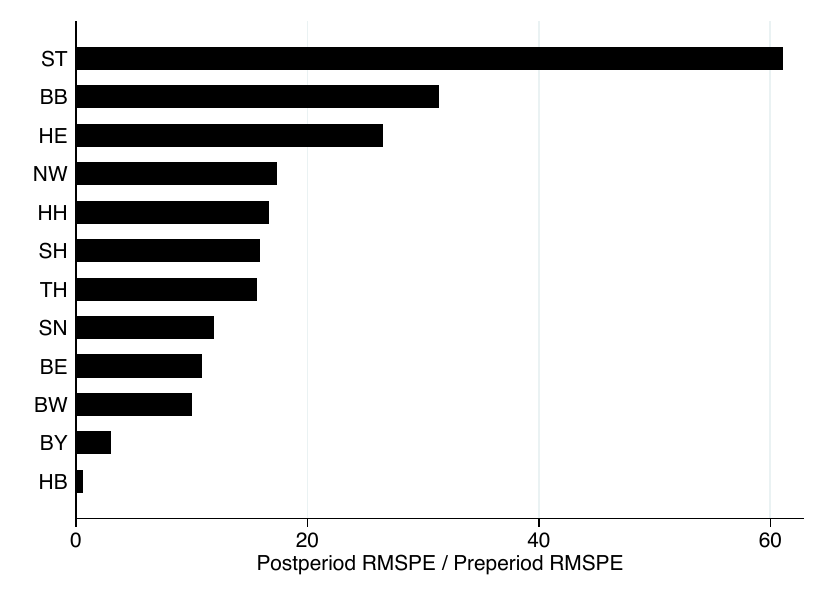}
	\end{subfigure}
	\begin{subfigure}[b]{\textwidth}
	\caption{Wind energy}
	    \includegraphics[width=0.325\textwidth,trim=0 0cm 0 0cm]{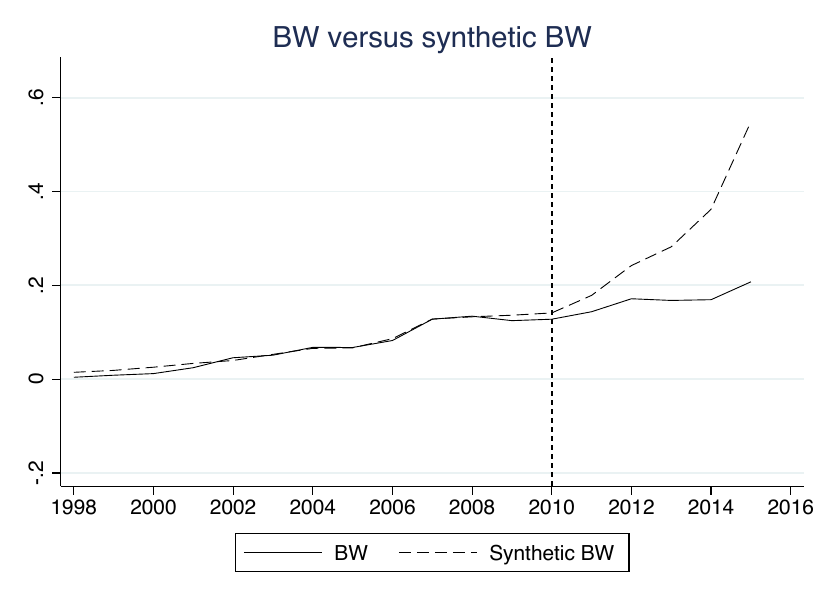}
		\includegraphics[width=0.325\textwidth,trim=0 0cm 0 0cm]{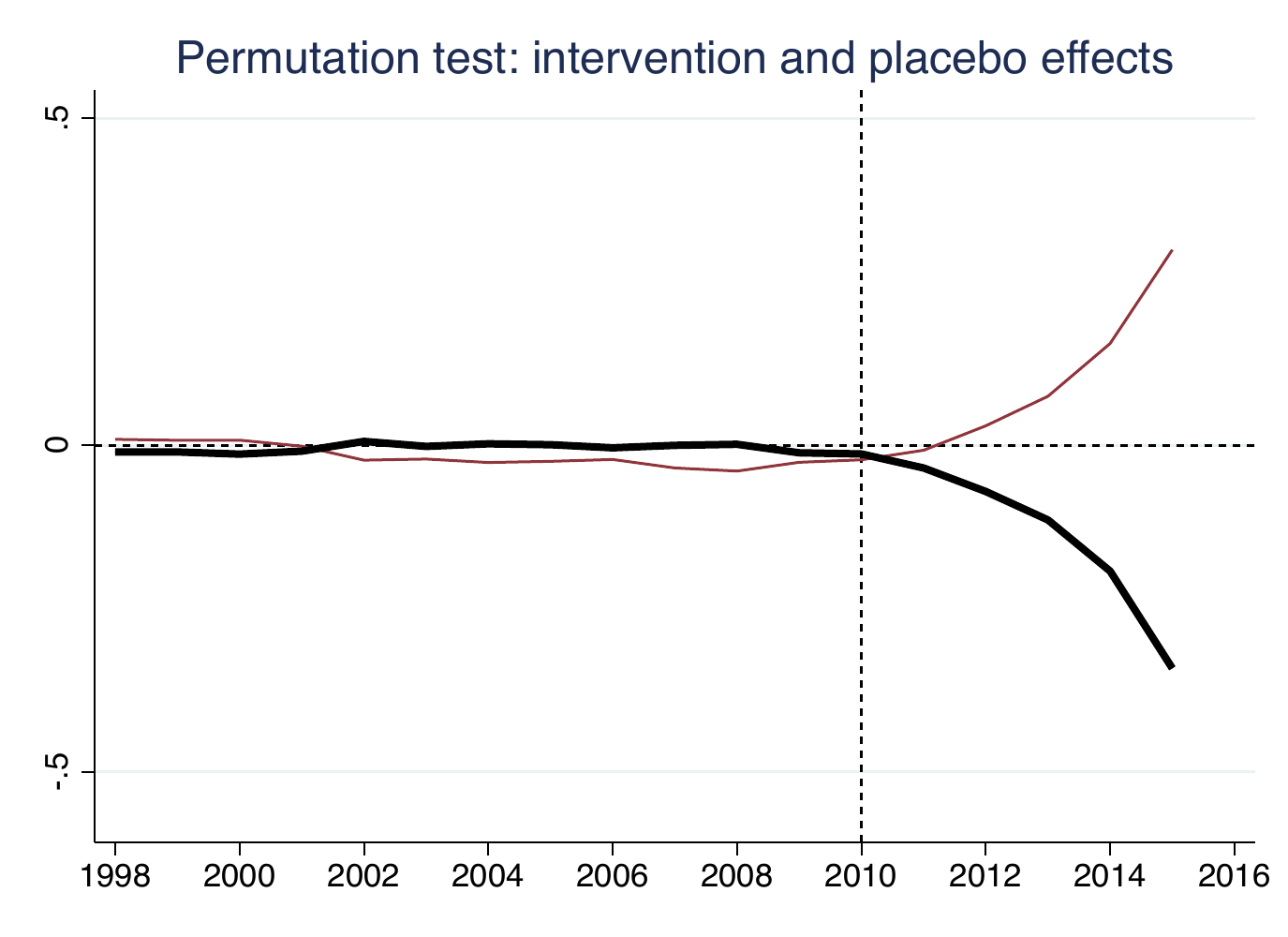}
		\includegraphics[width=0.325\textwidth,trim=0 0cm 0 0cm]{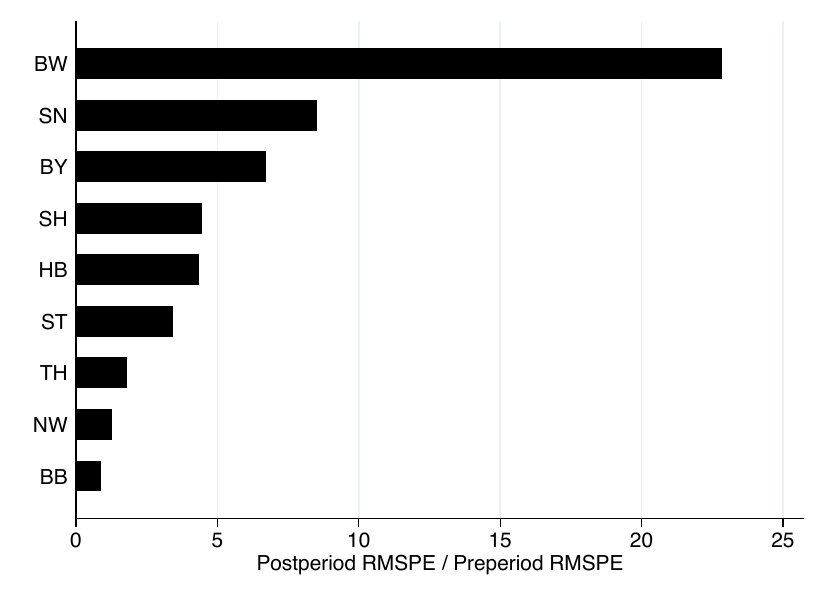}
	\end{subfigure}
		\vspace{0.1cm}
	\end{center}
	\footnotesize{\textit{Notes:} The center figures exclude states for which the pre-treatment MSPE is at least 10 times larger than BW's pre-treatment MSPE. The data are from the State Working Committee for Energy Balances.}
	\label{fig: Renewable engergies}
\end{figure}

Next, we investigate renewable energies. Figure \ref{fig: Renewable engergies} presents the results. Renewable energies as a share of primary energy usage were similar in BW and synthetic BW before treatment. The permutation inference procedure does not suggest that the Green government influenced the share of renewable energies. We also do not observe significant effects on the share of solar energy. The share of solar energy decreased under the Green government, but the RMSPE ratio for BW is not among the largest among all states.  

How the Green government influenced wind power is remarkable. Figure \ref{fig: Renewable engergies} suggests that the Green government significantly decreased wind power measured as a share of primary energy usage: the ratio of post- to pre-treatment RMSPE is the largest for BW. The effect is based on a larger increase in wind energy usage in the synthetic BW than in BW. Bavaria receives an SC weight of 94.8\% (cf.\ Table \ref{tab: tableenergyweights}). The Green government in BW promoted wind power to a smaller extent than the conservative CSU in Bavaria after the Fukushima disaster.

The negative effect on wind energy usage contrasts with expectations one may well have about Green governments. It does corroborate, however, anecdotal evidence about how the Green government influenced wind power in BW \citep[e.g.,][]{goetz2019dilemma}. During the election campaign before the elections in March 2011, the Green party promised to increase wind power and to promote direct democracy and public participation. For example, the Green government introduced a new Ministry of State for public participation. As a result, the Green government needed to handle citizens' action committees (``not-in-my-backyard'') and trade-offs between building wind turbines and preserving natural habitats for animals such as birds.\footnote{``Not-in-my-backyard'' movements emphasize that wind turbines give rise to direct costs for voters who wish to enjoy potential benefits from renewable energies. On voters' demand for ``bad policies'' see, for example, \citet{dal2018demand}.} Wind turbines curtail natural habitats for animals. First, they are life-threatening for birds and bats.\footnote{See, for example, \citet{serranoetal2020} for evidence from Spain.} In BW, nature conservation associations were opposing wind turbines because the wind turbines curtail the habitats of bats and red kites. Second, installing and maintaining wind turbines requires cutting down vegetation and curtailing habitats of animals that live on the ground. Thus, the Green government had to handle an intra-ecological conflict: nature and animal protection versus climate protection \citep{wurster2017energiewende}. Moreover, the Green state government did not enjoy encompassing political support at the local level. Having Green political majorities in the local and city councils may have helped to expand wind power as well (should the Green government have liked to do so given the intra-ecological conflict).

When the treatment (i.e., having a Green government) is randomly assigned, the permutation $p$-values reported in this section can be interpreted as classical randomization $p$-values for testing the null hypothesis that the Green government had no impact whatsoever \citep[e.g.,][]{abadie2010synthetic,firpo18synthetic,abadie2020jel}. However, random assignment may not be plausible in our context, for example, because the vote share of the Green party before 2011 likely affects the probability of getting a Green government. When the treatment is not randomly assigned, the permutation $p$-values can be interpreted as evaluating ``significance relative to a benchmark distribution for the assignment process'' \citep[][p.\ 404]{abadie2020jel}. Because this interpretation may not always be satisfactory, we employ the conformal inference procedure of \citet{chernozhukov2019exact} as a robustness check in Section \ref{sec: robustness}.
This method does not rely on random assignment but instead imposes weak dependence and stationarity of the SC prediction errors. Employing the conformal inference procedure does not change the inferences.

In Appendix \ref{app: add energy} we also investigate whether the Green government expanded protected nature reserves and landscape conservation areas (measured as a share of the state's overall area). We find no evidence of a significant effect of the Green government on either outcome.\footnote{In 2014, the Green government introduced BW's first national park: the Northern Black Forest. Because the Northern Black Forest is BW's only national park, and other German states either have no or only very few national parks, national parks are not a suitable outcome for SC methods.} 

In response to the Fukushima disaster, the federal government decided to close 6 out of the 17 nuclear power plants in 2011, including one nuclear power plant in BW. This decision may have had different effects on states with and without nuclear power plants. Therefore, in Appendix \ref{app: nuclear power plants}, we apply SC based on a restricted donor pool with the four states in which nuclear power plants were closed in 2011. Our results are robust to restricting the donor pool.

\subsection{Macroeconomic outcomes}
\label{sec: Macroeconomic outcomes}

Ex-ante, the effect of a Green government on macroeconomic outcomes is unclear.
On the one hand, the partisan theories \citep{hibbs1977political,chappell1986party,alesina1987macroeconomic} suggest that GDP and employment increase in the short-run when leftwing governments enter office, and Green parties have traditionally belonged to the leftwing political camp. An important reason is that leftwing governments are expected to implement expansionary policies such as increasing public expenditure. Indeed, the new Green government was advocating a larger size and scope of government than the previous CDU-led governments \citep{hoerisch2017editbook}. 
On the other hand, the partisan theories were developed for traditional party systems in the 1970s and 1980s, ignoring Green governments.  
Moreover, there had not been any other Green state (or national) government in Germany before. As a result, political uncertainty was pronounced, and citizens and entrepreneurs might have been hesitant in consuming and investing \citep[e.g.,][]{julio2012uncertainty}, which would decrease GDP in the short-run.

Our outcomes of interest are GDP per employee in Euros and the unemployment rate. The data are provided by the Research Group on Regional Accounts from the Federal Statistical Office and the Federal Agency of Work. 

\begin{figure}[H]
    \caption{Spaghetti graphs: macroeconomic outcomes}
        \vspace{-0.4cm}

    \begin{center}
    \begin{subfigure}[b]{0.325\textwidth}
         \centering
         \caption{GDP per employee}
         \includegraphics[width=\textwidth]{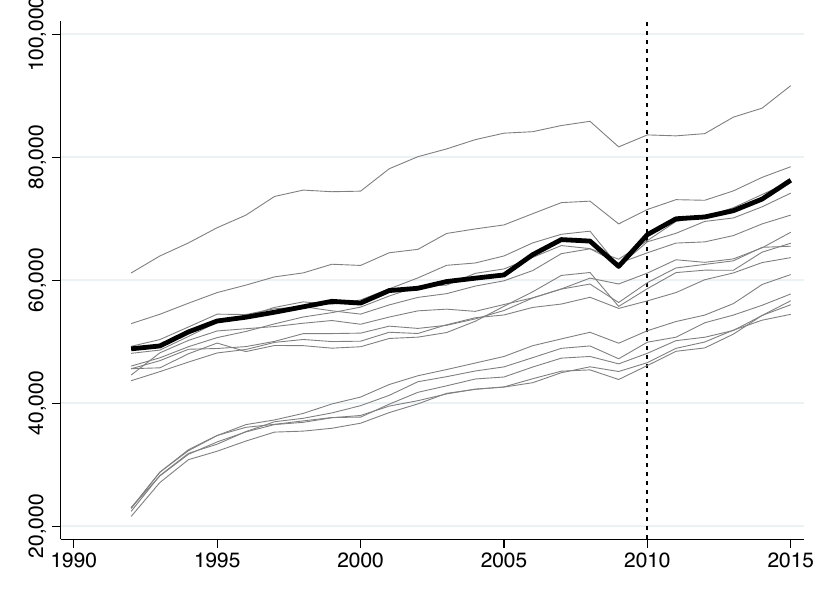}
    \end{subfigure}
    \begin{subfigure}[b]{0.325\textwidth}
        \centering
        \caption{Unemployment rate}
        \includegraphics[width=\textwidth]{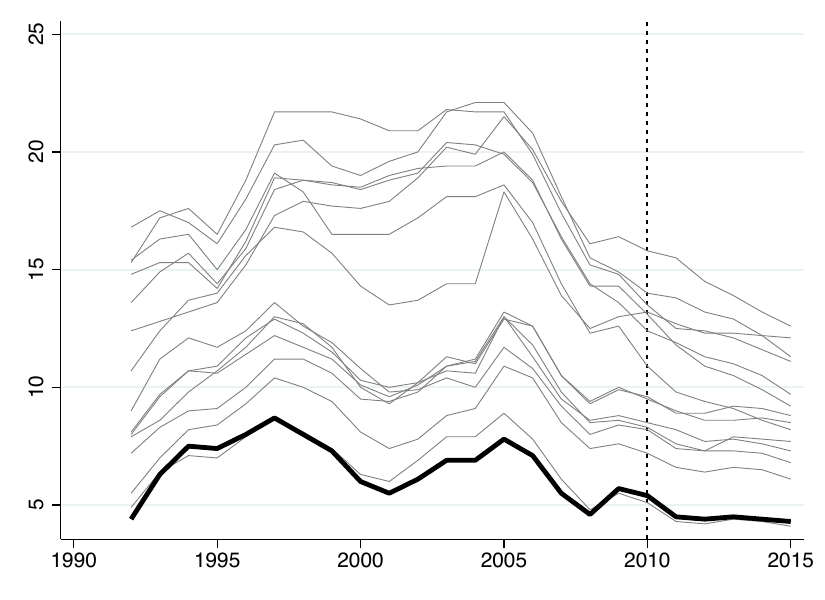}
    \end{subfigure}    
    \end{center}     
         \footnotesize{\textit{Notes:} The data are form the Regional Accounts VGRdL and Federal Agency of Work.}
\label{fig:head_med_macro}
\end{figure}

We first investigate how both macroeconomic outcomes developed in BW compared to the other German states. Figure \ref{fig:head_med_macro} shows that the levels of macroeconomic outcomes in BW (thick line) are different from the other German states. GDP per employee in BW is higher, while the unemployment rate in BW is lower than in most other German states. Table \ref{tab: tablemacroweights} in Appendix \ref{app: sc weights} shows the SC weights for the macroeconomic outcomes.

\begin{figure}[H]
	\caption{SC results: macro outcomes}
	    \begin{subfigure}[b]{\textwidth}
         \centering
         \caption{GDP per employee}
	\begin{center}
		\includegraphics[width=0.325\textwidth,trim=0 0 0 1cm]{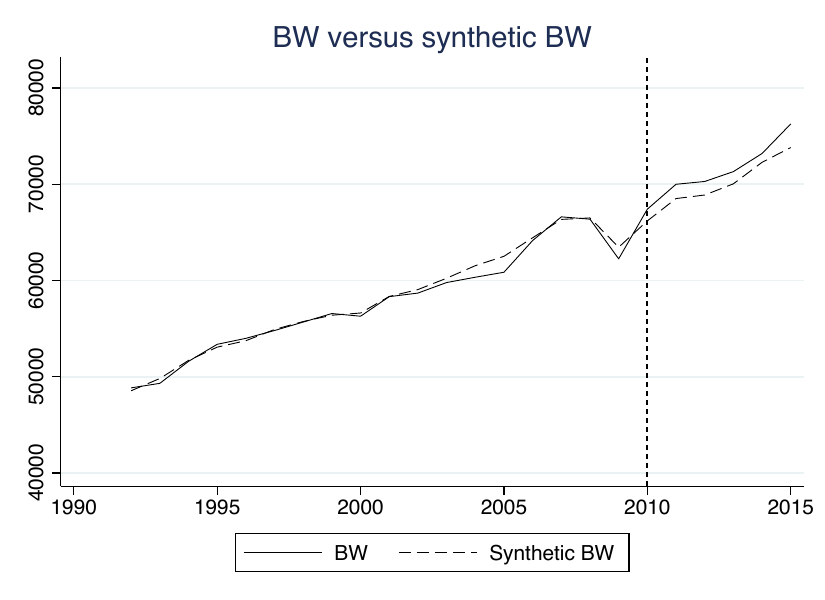}
        \includegraphics[width=0.325\textwidth,trim=0 0cm 0 1cm]{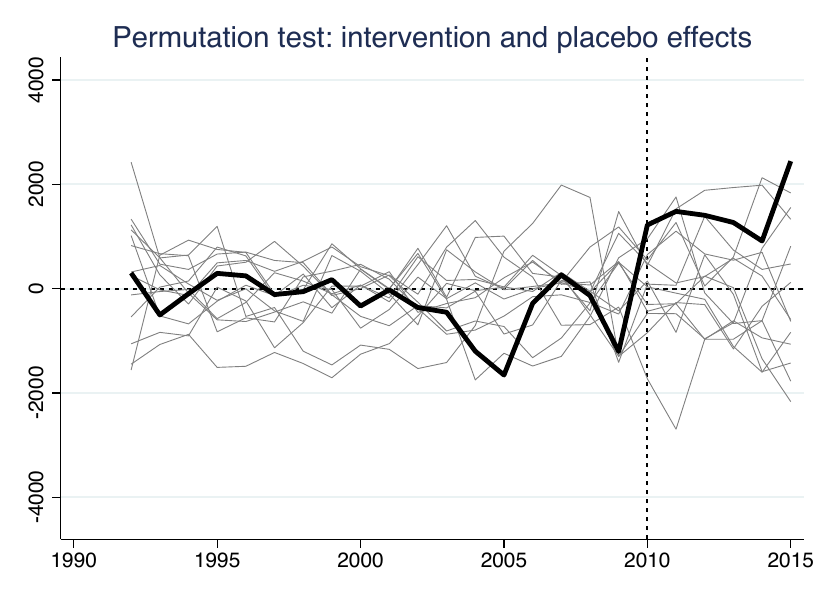}
        \includegraphics[width=0.325\textwidth,trim=0 0cm 0 1cm]{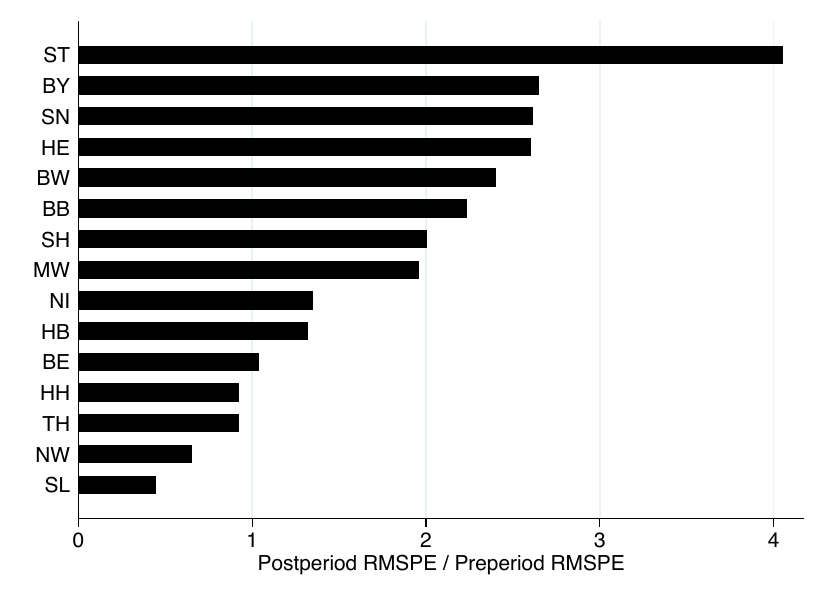}
	\end{center}
	\vspace{0.1cm}
	\end{subfigure}
		    \begin{subfigure}[b]{\textwidth}
         \centering
	\caption{Unemployment rate}
	\begin{center}
		\includegraphics[width=0.325\textwidth,trim=0 0 0 1cm]{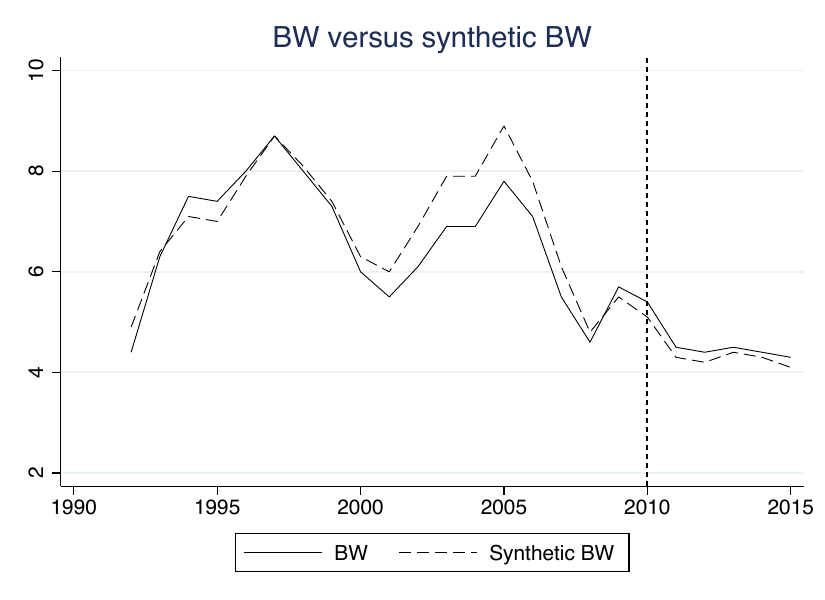}
        \includegraphics[width=0.325\textwidth,trim=0 0 0 1cm]{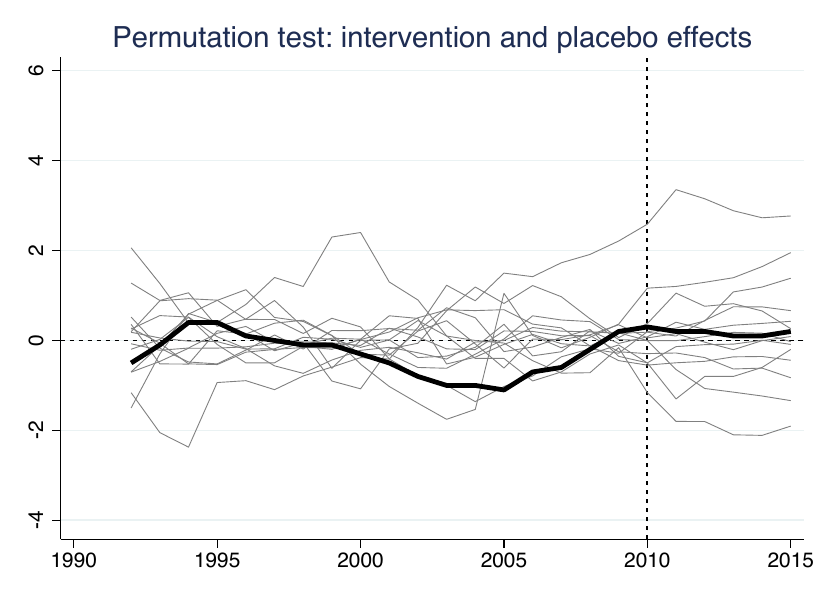}
        \includegraphics[width=0.325\textwidth,trim=0 0 0 1cm]{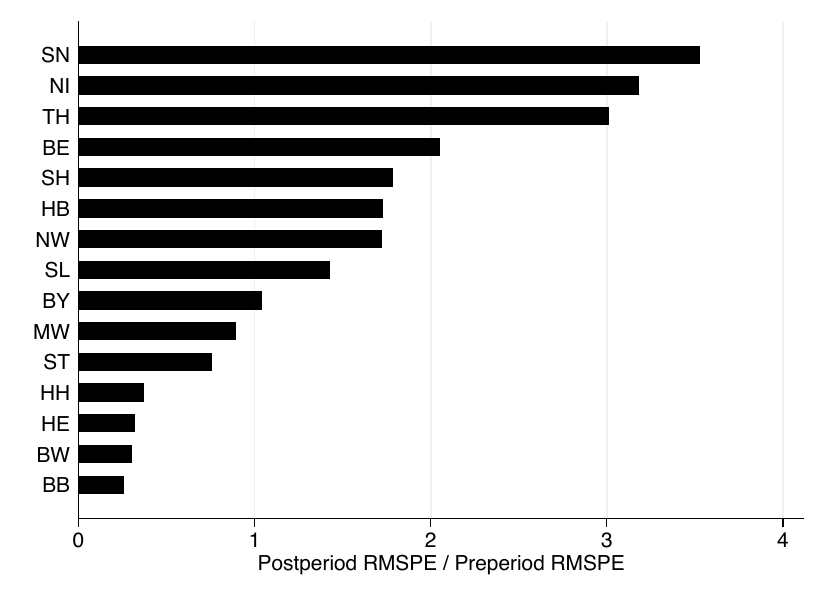}
	\end{center}
	\end{subfigure}
	\footnotesize{\textit{Notes:} The center figures exclude states for which the pre-treatment MSPE is at least 10 times larger than BW's pre-treatment MSPE. The data are from Regional Accounts VGRdL and the Federal Agency of Work.}
	\label{fig: macro outcomes}
\end{figure}

Figure \ref{fig: macro outcomes} presents the SC results. We find that GDP per employee was somewhat higher in BW than in synthetic BW after treatment, whereas the unemployment rate was hardly higher in BW than in synthetic BW after treatment. For GDP per employee, the ratio for BW is the fifth largest among all states. The implied permutation $p$-value is $p = 5/15=1/3$. Thus, the results do not suggest that the Green government increased GDP per employee. For the unemployment rate, we do not find a significant effect of the Green government: the RMSPE ratio for BW is the second smallest among all states.

\subsection{Education outcomes}

German state governments enjoy quite some leeway in designing education policies. A major issue is how state governments organize the school system. Ten years of school are mandatory in Germany. Students attend primary school for four years in BW, and must attend secondary school at least for another six years. There were four types of secondary schools: lower secondary schools (\textit{Hauptschulen}), secondary schools (\textit{Realschulen}), high schools (\textit{Gymnansien}), and comprehensive schools (\textit{Gesamtschulen}). Students attend high school for eight or nine years (instead of six years in lower secondary and secondary schools) and receive their high school diploma (\textit{Abitur}) when they graduate. In comprehensive schools, students with varying abilities attend the same school but are taught in individual classes. There are usually three performance levels in comprehensive schools, corresponding to the performance levels in lower secondary school, secondary school, and high school.

Before the 2011 election, the Greens in BW advocated collective studying for ten years in an elementary or comprehensive school. They emphasized that communities and society in general should have more power in designing school structures \citep{busemeyerhaaster2017bildungspolitik}. After taking office, the Green government introduced so-called community schools (\textit{Gemeinschaftsschulen}).
In community schools, students with different abilities are taught in the same class. 

Secondary schools were free to choose whether they wanted to become a community school; the state government provided financial incentives to do so. While many lower secondary schools and secondary schools became community schools, much fewer high schools were interested in becoming a community school \citep{faq2020}. The number of community schools was growing fast. In the school year 2012/2013, there was a starting group 42 community schools. One year later, there were already 89 community schools.

We examine how the Green government influenced the organization of secondary schools. We estimate SC models for the number of students in each individual school type: lower secondary schools, secondary schools, high schools, and comprehensive schools (including community schools). The data and classification are provided by the Standing Conference of the Ministers of Education and Cultural Affairs of the Laender (CMC).\footnote{The CMC subsumes the community schools to be comprehensive schools.} 
We discuss results for high schools and comprehensive schools here and the  results for lower secondary schools and secondary schools in Appendix \ref{app: add education}. Table \ref{tab: tableeducationweights} in Appendix \ref{app: sc weights} shows the SC weights for the education outcomes.

\begin{figure}[ht]
    \caption{Spaghetti graphs: number of students}
    \begin{center}
        \vspace{-0.4cm}
    \begin{subfigure}[b]{0.325\textwidth}
         \centering
         \caption{Compr. schools}
         \includegraphics[width=\textwidth]{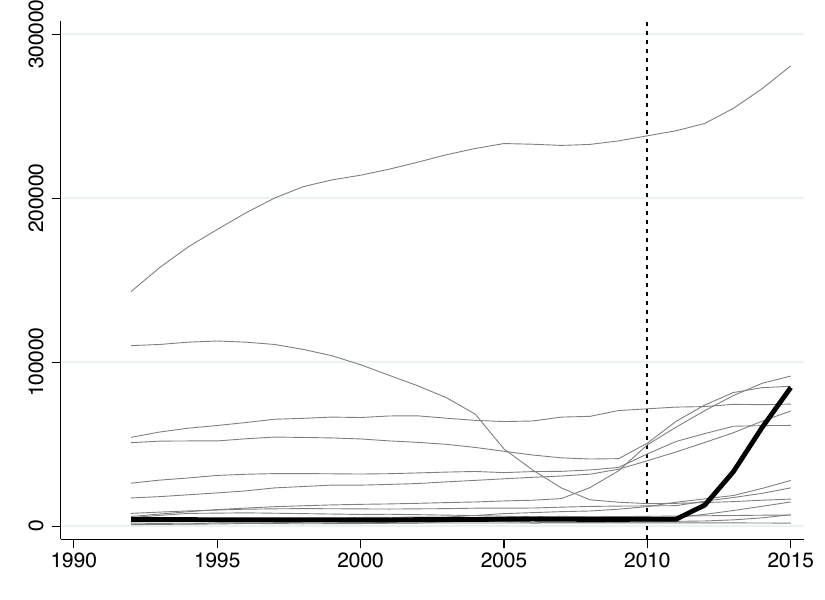}
    \end{subfigure}
    \begin{subfigure}[b]{0.325\textwidth}
         \centering
         \caption{High schools}
         \includegraphics[width=\textwidth]{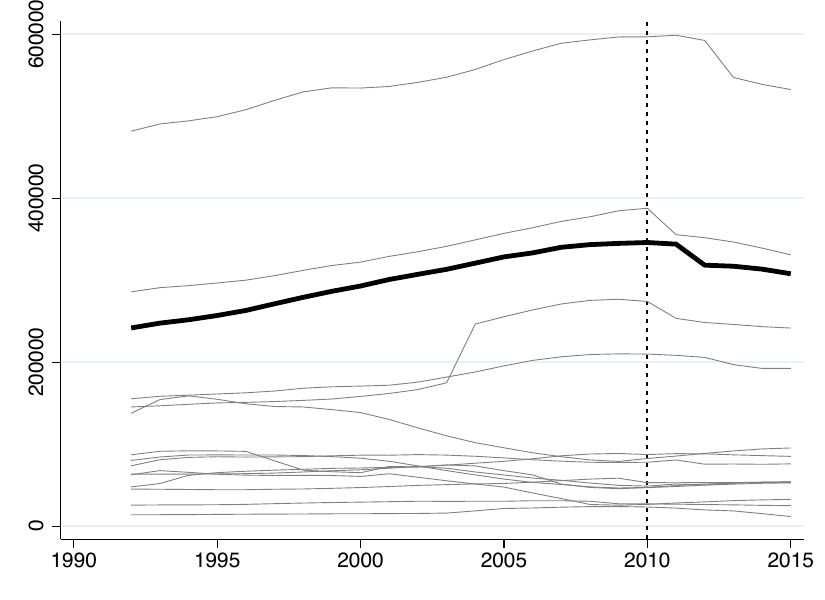}
    \end{subfigure}
    \end{center}
 \footnotesize{\textit{Notes:} The data are from the Standing Conference of the Ministers of Education and Cultural Affairs of the Laender.}       
\label{fig:spaghetti_schools}
\end{figure}

\begin{figure}[H]
	\caption{SC results: number of students}
	    \begin{subfigure}[b]{\textwidth}
         \centering
         \caption{Number of students in comprehensive schools}
	\begin{center}
		\includegraphics[width=0.325\textwidth,trim=0 0cm 0 1cm]{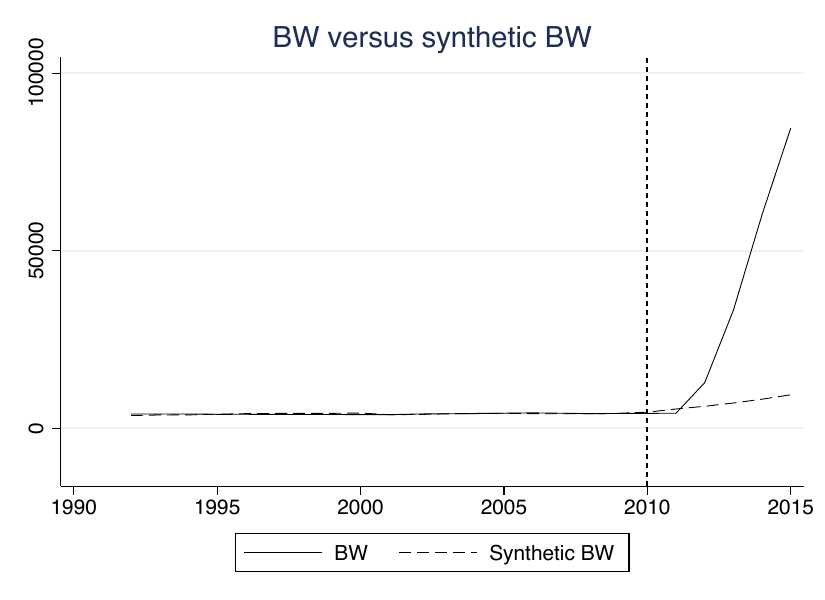}
        \includegraphics[width=0.325\textwidth,trim=0 0cm 0 1cm]{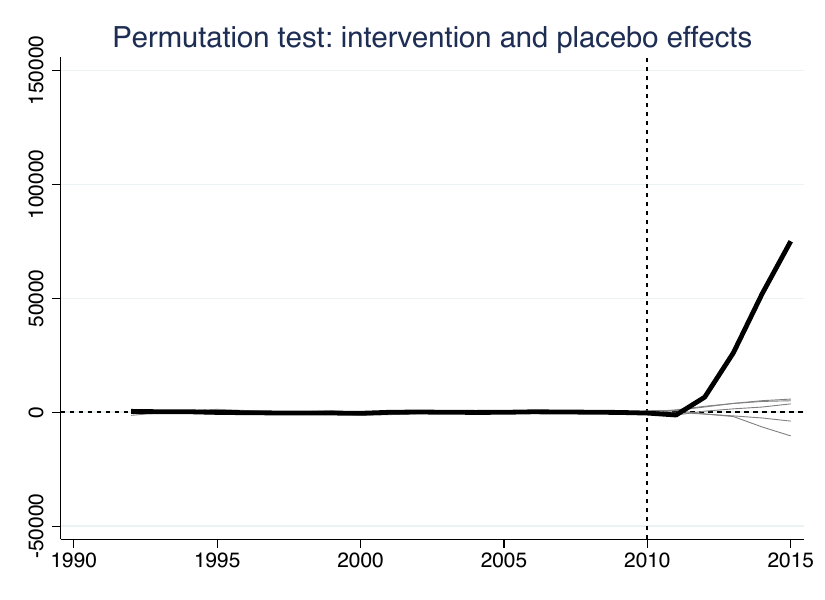}
        \includegraphics[width=0.325\textwidth,trim=0 0cm 0 1cm]{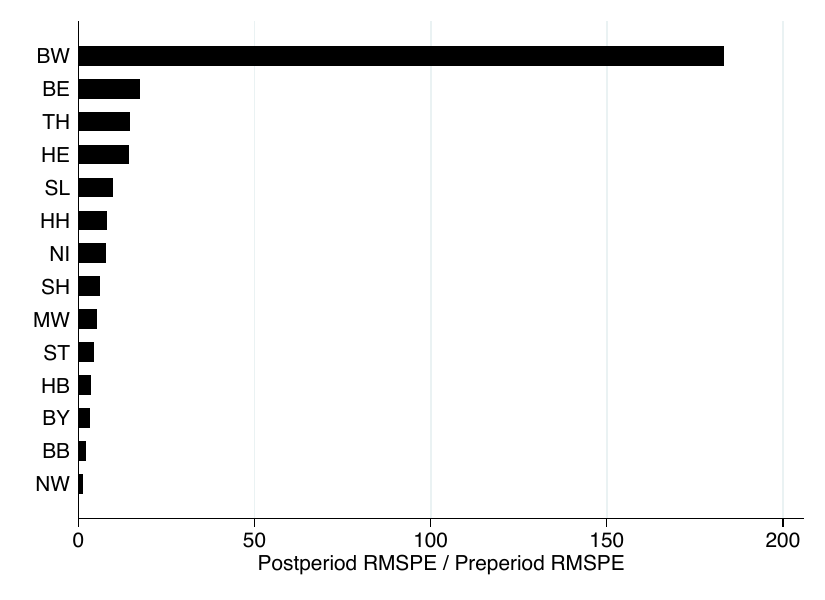} 
	\end{center}
	\vspace{0.1cm}
	\end{subfigure}
		    \begin{subfigure}[b]{\textwidth}
         \centering
	\caption{Number of students in high schools}
	\begin{center}
		\includegraphics[width=0.325\textwidth,trim=0 0cm 0 1cm]{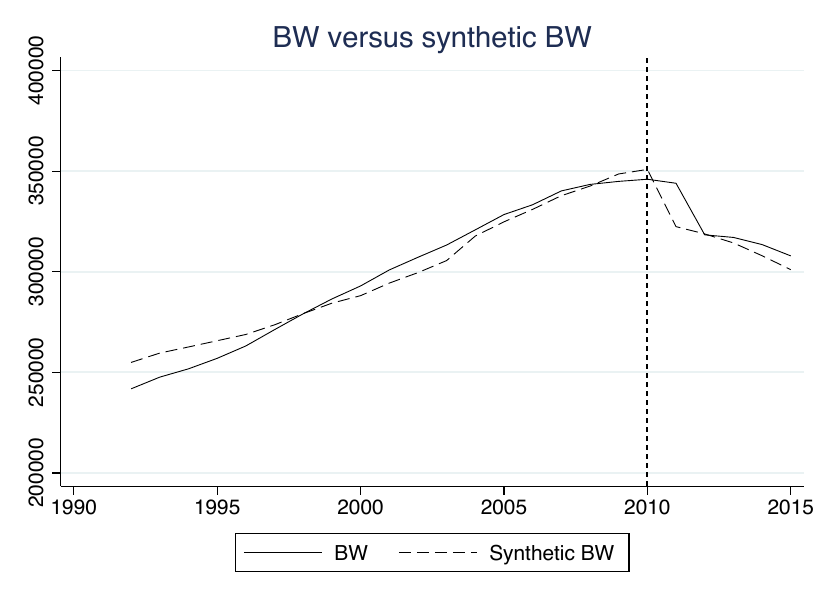}
        \includegraphics[width=0.325\textwidth,trim=0 0cm 0 1cm]{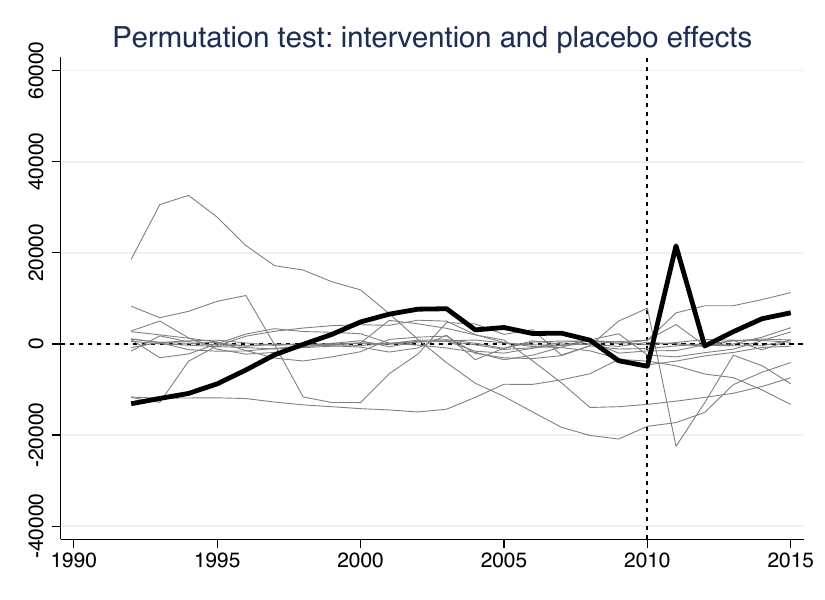}
        \includegraphics[width=0.325\textwidth,trim=0 0cm 0 1cm]{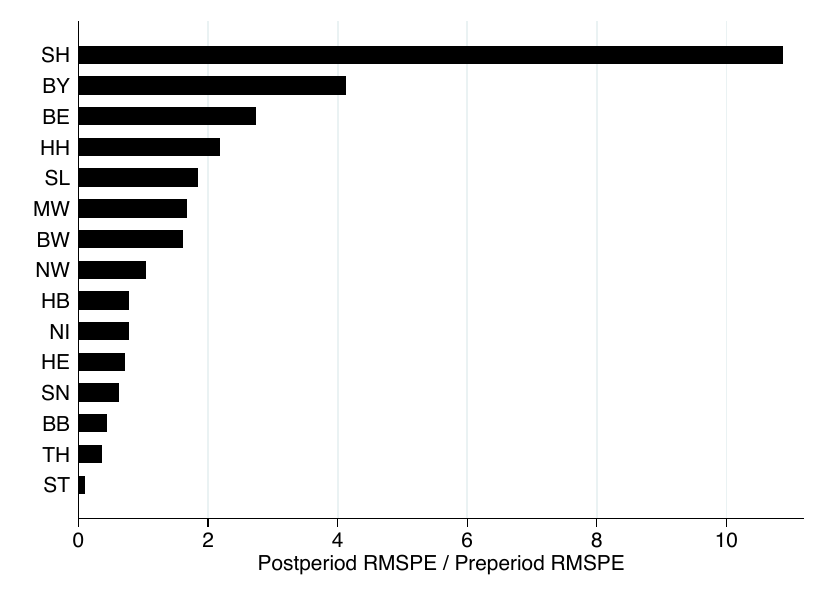}
	\end{center}
	\end{subfigure}
	 \footnotesize{\textit{Notes:} The center figures exclude states for which the pre-treatment MSPE is at least 10 times larger than BW's pre-treatment MSPE. The data are from the Standing Conference of the Ministers of Education and Cultural Affairs of the Laender.}
	\label{fig: Schools}
\end{figure}

Figure \ref{fig: Schools} shows that the SC method delivers an excellent pre-treatment fit for students at comprehensive schools. The gaps between students in BW and the synthetic BW after treatment are huge. The permutation inference procedure suggests that these gaps are the largest among the German states (center panel). The ratio of post- to pre-treatment RMSPE is by far the largest in BW compared to the other German states (right panel) and  the implied permutation $p$-value is $1/14\approx 0.07$ (the smallest possible $p$-value). These results suggest that the change of government caused a substantial expansion of comprehensive schools, an effect based on introducing community schools.\footnote{The results for comprehensive schools are robust to standardizing the number of students by the overall population and by the total number of students in each state.} 

The pre-treatment fit for students at high schools is somewhat worse than the pre-treatment fit for students at comprehensive schools. The results suggest that the Green government did not influence the number of students in high schools.

\section{Robustness checks}
\label{sec: robustness}

This section presents the findings of two standard robustness checks and reports the results from applying the conformal inference method of \citet{chernozhukov2019exact}. We focus on the two outcomes where the RMSPE ratio for BW is the largest among all states:  the share of wind energy and the number of students in comprehensive schools.

\begin{figure}[H]
	\caption{Placebo treatment in 2006}
		\begin{subfigure}[b]{\textwidth}
	\caption{Wind energy}
	     	\begin{center}
		\includegraphics[width=0.325\textwidth,trim=0 0cm 0 1cm]{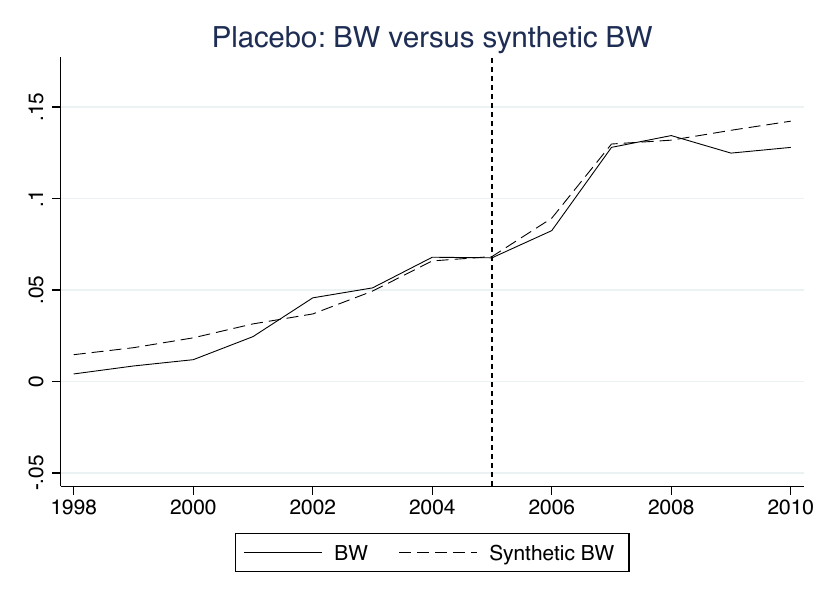}
		\includegraphics[width=0.325\textwidth,trim=0 0cm 0 1cm]{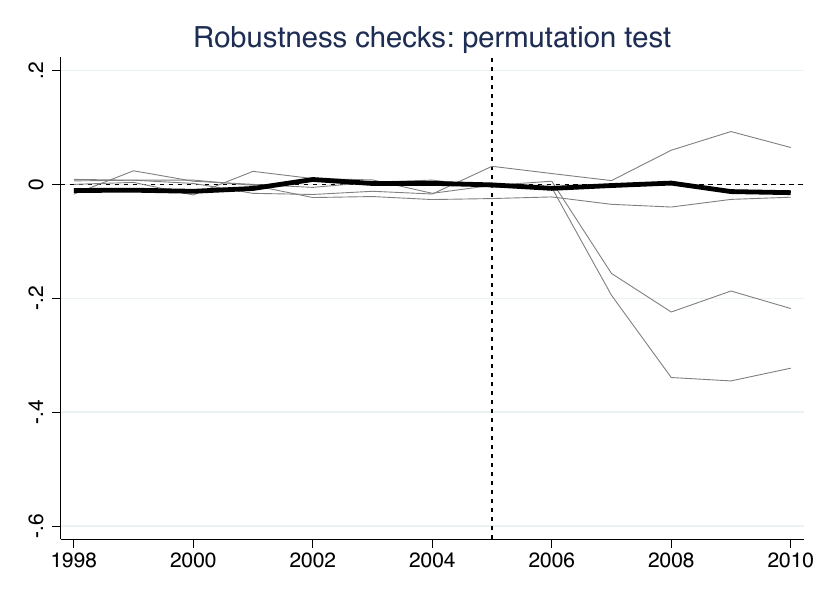}
		\includegraphics[width=0.325\textwidth,trim=0 0cm 0 1cm]{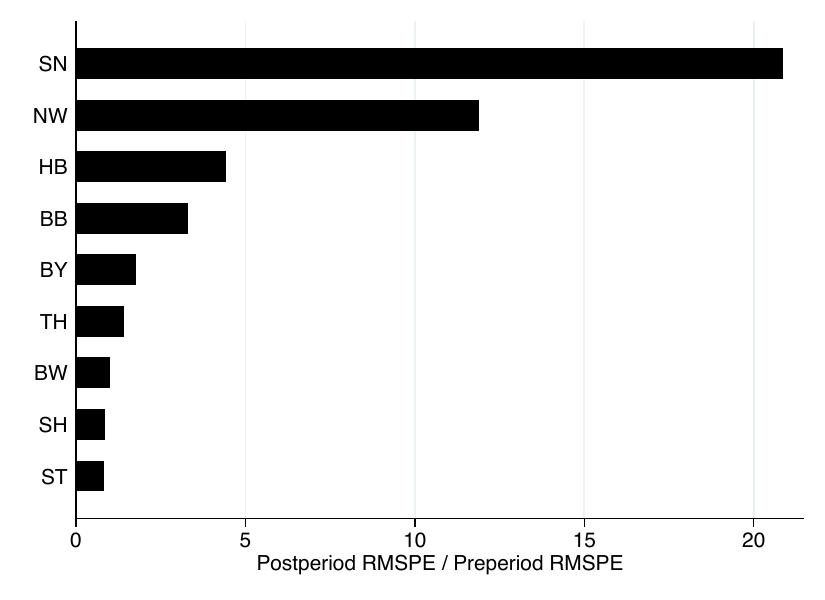}
	\end{center}
	\end{subfigure}
	
	\begin{subfigure}[b]{\textwidth}
	\caption{Students comprehensive schools}
	\begin{center}
		\includegraphics[width=0.325\textwidth,trim=0 0cm 0 1cm]{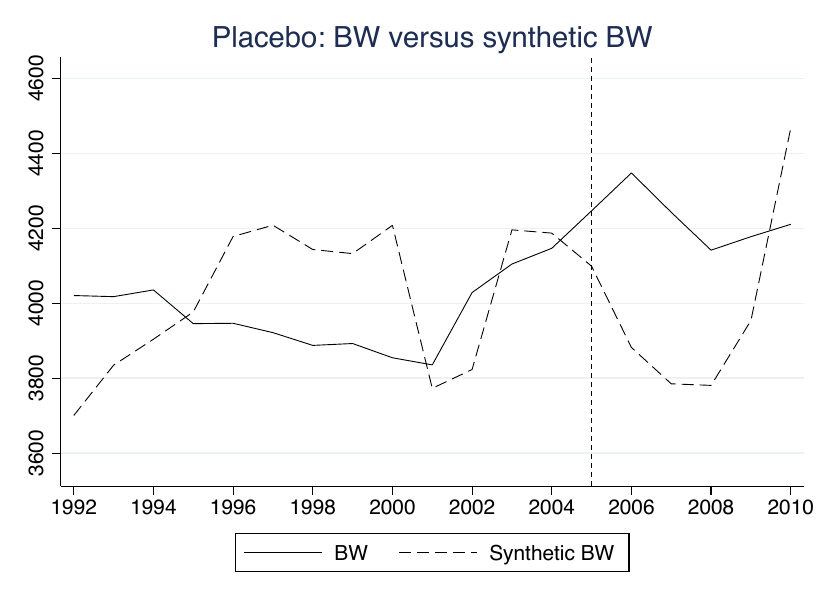}
        \includegraphics[width=0.325\textwidth,trim=0 0cm 0 1cm]{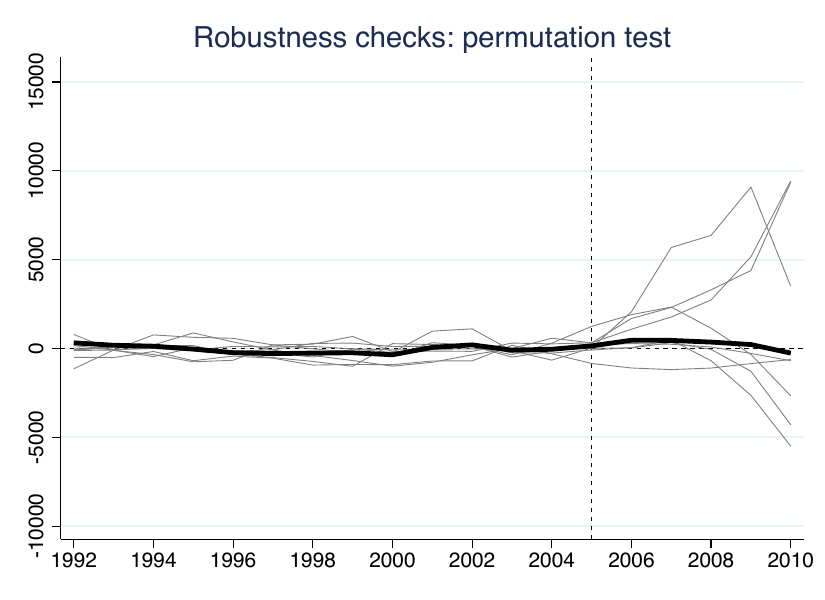}
        \includegraphics[width=0.325\textwidth,trim=0 0cm 0 1cm]{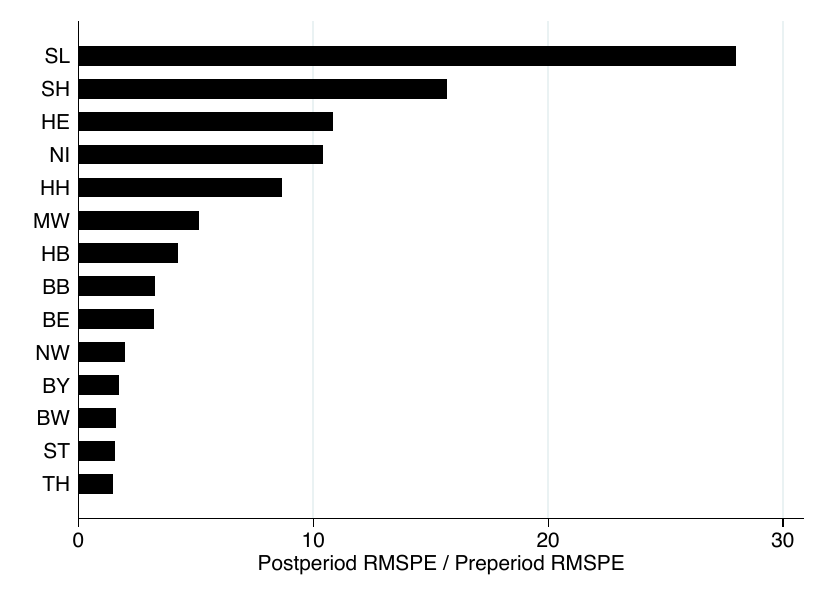}        
	\end{center}
	\end{subfigure}
	
	\vspace{0.1cm}

	\footnotesize{\textit{Notes:} The center figures exclude states for which the pre-treatment MSPE is at least 10 times larger than BW's pre-treatment MSPE. The data are from the State Working Committee for Energy Balances and from the Standing Conference of the Ministers of Education and Cultural Affairs of the Laender.}
	\label{fig: Placebo treatment}
\end{figure}

We start by performing ``in-time'' placebo tests as suggested by \citet{abadie2015comparative}. In particular, we backdate the treatment from 2011 to 2006 (the year of the previous election). Figure \ref{fig: Placebo treatment} shows that there are no significant placebo effects, which corroborates our main findings. 

As a second robustness check, we perform ``leave-one-out'' sensitivity analyses to examine whether our results are driven by an influential control state \citep[e.g.,][]{abadie2015comparative}. We iteratively exclude from the donor pool individual states that received positive weight in our SC analysis and apply SC to the restricted donor pool.

\begin{figure}[H]
    \caption{Leave-one-out sensitivity}
    \vspace{-0.4cm}
    \begin{center}
    \begin{subfigure}[b]{0.325\textwidth}
        \centering
        \caption{Wind energy}
        \includegraphics[width=\textwidth,trim=0 0 0 0.5cm]{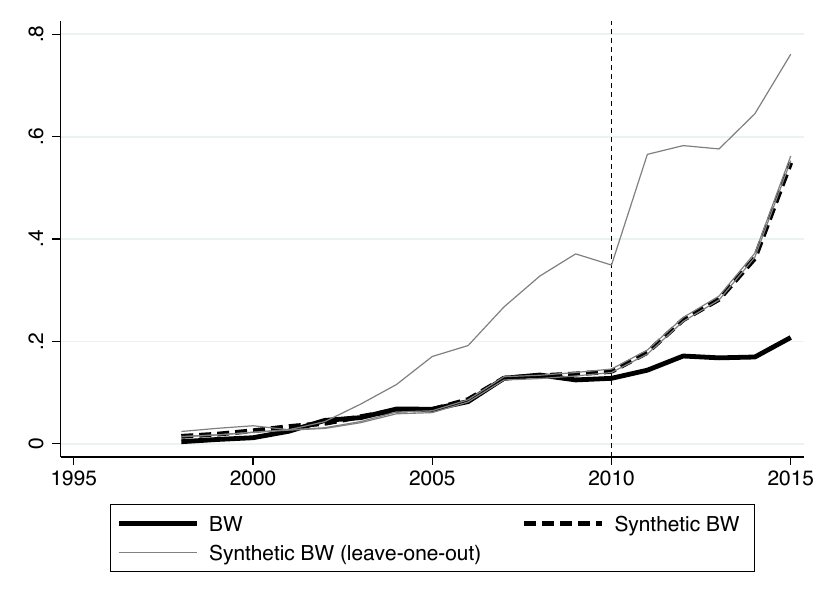}
    \end{subfigure}
    \begin{subfigure}[b]{0.325\textwidth}
         \centering
         \caption{Students comp. schools}
         \includegraphics[width=\textwidth,trim=0 0 0 0.5cm]{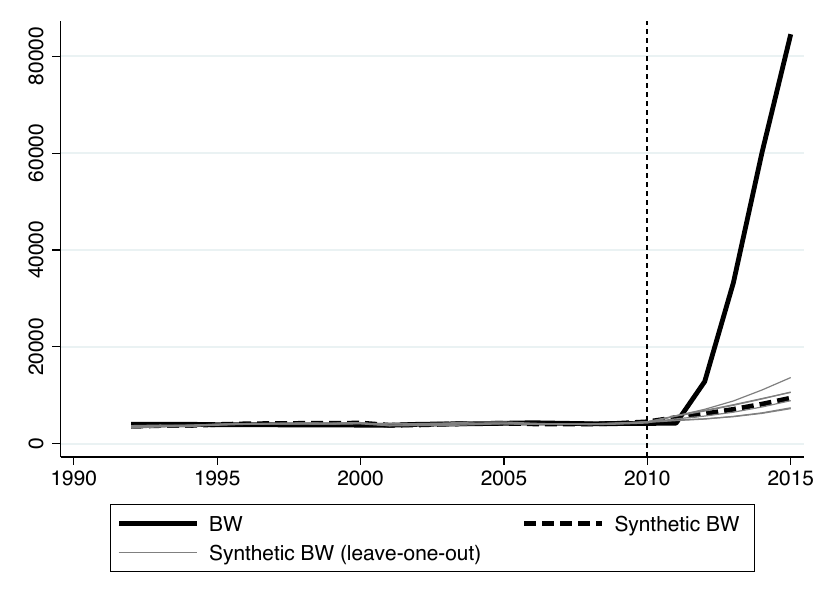}
    \end{subfigure}

\end{center}

    \footnotesize{\textit{Notes:} The data are from the State Working Committee for Energy Balances and from the Standing Conference of the Ministers of Education and Cultural Affairs of the Laender.}
\label{fig: LOO}
\end{figure}

Figure \ref{fig: LOO} presents the results. The results for wind energy depend on considering Bavaria as a control unit (Panel (a)). Bavaria receives 95\% of the weight in the synthetic BW for wind energy usage. When we exclude Bavaria, the estimated effect gets larger, while the pre-treatment fit becomes worse. This result is expected: BW and Bavaria were very similar before the treatment, both with respect to the level and the trend, and no other control state provides a good fit. It demonstrates that including Bavaria, which is a ``natural'' control unit (cf.\ Section \ref{subsec: choice of donor pool}), into our donor pool is important for our empirical strategy. The findings for students at comprehensive schools are essentially invariant to excluding individual control states (Panel (b)).

Finally, we employ the inference method proposed by \citet{chernozhukov2019exact}. The idea of this procedure is to compare the SC residuals in the post-treatment period to those in the pre-treatment period. ``Large'' post-treatment residuals provide evidence against the null hypothesis that the Green government had no impact whatsoever. \citet{chernozhukov2019exact} operationalize this idea by proposing a conformal inference method based on permuting SC residuals. We compute $p$-values using both moving block and iid permutations.\footnote{See Section 2.2 in \citet{chernozhukov2019exact} for a further discussion of these two types of permutations. To ensure computational tractability, we follow \citet{chernozhukov2019exact} and compute $p$-values based on 10,000 iid permutations.} The $p$-values based on the moving block permutations are valid if the SC prediction errors are stationary and weakly dependent. The iid permutations allow for computing more precise $p$-values but only yield valid inferences with iid prediction errors.

For the share of wind energy, the $p$-values are 0.111 (moving block permutations) which is equal to the corresponding $p$-value based on the permutation approach, and 0.006 (iid permutations). The $p$-value for students in comprehensive schools based on the moving block permutations is 0.042; the $p$-value based on the iid permutations is 0.001. Thus, the conformal inference results confirm the inferences based on the \citet{abadie2010synthetic} permutation procedure.

\section{Conclusion}

Combating climate change requires fundamental policy changes. Green parties are expected and promise to improve environmental outcomes. Over the last decades, they have enjoyed electoral success and have gained more and more executive power. Therefore, a major question is what Green parties do when they are in office and lead governments.

To investigate how Green governments influence environmental and economic outcomes, we exploit that the Fukushima nuclear disaster in Japan gave rise to an unanticipated change of government in the German state BW. The Green party benefited from the Fukushima nuclear disaster and has set the prime minister in BW since 2011; the first and so far only Green prime minister in Germany. To estimate causal effects, we employ the SC method which provides a transparent and data-driven approach for choosing a suitable control group for BW.

We do not find evidence that the Green government influenced macroeconomic outcomes such as GDP per employee and the unemployment rate. The Green government implemented education policies, promoting more inclusive schools. Our results suggest that these policies caused the number of students in comprehensive schools (including community schools) to increase. 

A key result of our study is that environmental and energy outcomes did not change as one would expect. The Green government did not influence CO2 emissions or increase energy usage from renewable energies overall. It even decreased wind energy power compared to the estimated counterfactual. Against the background that the Fukushima nuclear disaster opened a window of opportunity for changes in environmental and energy outcomes, our results are stark.
It is likely that handling intra-ecological conflicts prevented the Green government from implementing drastic changes in environmental and energy outcomes. Expanding wind power, for example, gives rise to trade-offs. On the one hand, wind power is an alternative energy that decreases the relative energy usage of fossil fuels. On the other hand, wind turbines disfigure the landscape (the Greens in BW encouraged direct democracy and needed to deal with ``not in my back yard'' movements), and wind turbines curtail natural habitats of animals such as birds. Intra-ecological conflicts were pronounced, and the Green state government needed to handle those conflicts.

The Greens did not enjoy broad political majorities in the counties and municipalities. Some political projects such as expanding wind energy benefit from support at the local level. The lack of political support across all levels of governments made it difficult for the Green state government in BW to implement more policies that reflect their platforms. 

The Green government faced the same issues when being in office as any other government, including handling veto powers and re-election concerns. Such issues may prevent Green governments from implementing environmentally friendly policies and improving environmental outcomes. On the other hand, our results for wind energy suggest that political parties for which environmental policies have not been frontline issues, such as the CSU in Bavaria, may implement policies to improve environmental outcomes. Indeed, substantial policy changes are sometimes implemented by ``unlikely'' parties \citep{cukiermantommasi1998}.

\newpage
\bibliographystyle{apalike}
\interlinepenalty 10000
\bibliography{fukushima_bibliography.bib}

\newpage

\begin{appendix}

\setcounter{page}{1} 

\begin{center}
    \huge{Appendix to \emph{Green governments}}
    
    \vspace{0.5cm}
    
    \large{Niklas Potrafke \qquad Kaspar W\"uthrich}
\end{center}

\startcontents[sections]
\printcontents[sections]{l}{1}{\setcounter{tocdepth}{1}}

\section{Party competition in BW and the other German states}
\label{app: party competition}
One may wonder whether our results are driven by the change in government in 2011 being small because the ideological differences between the previous conservative and the new Green government were small. For example, the Greens in BW also won the 2016 state election and have formed a coalition with the conservative CDU since 2016. Therefore, we portray the individual positions of the political parties in the German states. \citet{braeuninger2020book} measure party platforms of the individual parties in the German states.\footnote{We are grateful to the authors for sharing their data with us.} They distinguish between economic policy positions (e.g., taxation and regulation) and social policies (e.g., immigration and homosexuality). Environmental policy positions are not considered because environmental policy positions have not been included in every manifesto that \citet{braeuninger2020book} examine over the period 1990--2019.

The positions are based on computer-assisted text analyses of manifestos. Computer programs search for individual keywords describing policy positions and count how often they are used and consider contexts. \citet{braeuninger2020book} obtain standardized scores that measure party positions for economic and social policies. Small values of the social policy measure mean that the parties prefer liberal social policies (e.g., advocating immigration and same-sex marriage), large values mean that the parties prefer conservative social policies (e.g., not advocating immigration and promoting traditional family values). In a similar vein, small values of the economic policy measure mean that the parties advocate a large size and scope of government, large values mean that the parties advocate market-oriented economic policies. 

The social policy positions in our sample assume values between -1.27 (the left party in North Rhine-Westphalia) and 15.13 (the CDU in Saxony-Anhalt). The mean and standard deviation are 7.17 and 3.33. The economic policy positions in our sample take values between -0.04 (the left party in North Rhine-Westphalia) and 18.45 (the FDP in Saxony-Anhalt). The mean and standard deviation are 11.23 and 3.74. Our sample covers 2008--2011 and includes one observation per party and state. These data are based on manifestos for state elections that take place every four to five years. We focus on the parties' positions in the year 2011 (when the Green government in BW took office) or earlier if state elections took place in 2010, 2009 or 2008.

We examine party positions of the five major political parties in the year 2011: the Left party, Greens, SPD, CDU, and FDP. The populist rightwing ``Alternative for Deutschland'' (AfD) was founded in the year 2013. 

\begin{figure}
	\caption{Social policy positions of political parties, state level, 2008--2011}
	\begin{center}
	\includegraphics[width=0.7\textwidth]{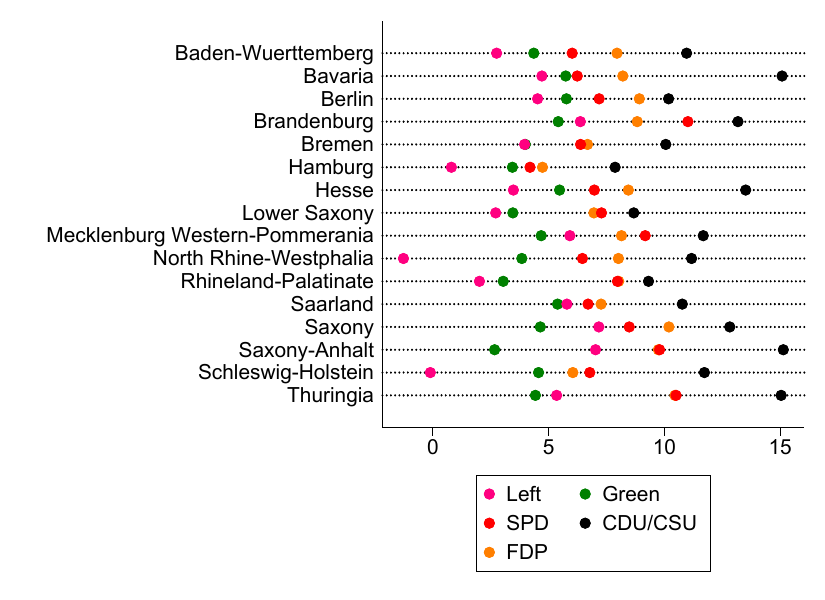}
	\end{center}
	\footnotesize{\textit{Notes:} The figure shows how social policy positions differ between the political parties in every individual state. Standardized scores measure party positions for social policies. Small values of the social policy measure mean that the parties prefer liberal social policies (e.g., advocating immigration and same-sex marriage), large values mean that the parties refer conservative social policies (e.g., not advocating immigration and promoting traditional family values). The data are from Appendix B in  \citet{braeuninger2020book}.}
	\label{fig: social policy positions}
\end{figure}

Figure \ref{fig: social policy positions} shows that social policy positions differ quite a bit between the political parties in every individual state. In BW, the Greens scored 4.36, and the CDU scored 10.96. These scores are quite comparable, for example, with Schleswig-Holstein, where the Greens scored 4.57, and the CDU scored 11.73. The Greens were somewhat more conservative in Bavaria (5.74), Berlin (5.77), and the Saarland (5.39) than in BW (4.36). The CDU was quite a bit more conservative in Bavaria (15.08), Saxony-Anhalt (15.13), and Thuringia (15.04) than in BW (10.96).

\begin{figure}[H]
	\caption{Economic policy positions of political parties, state level, 2008--2011}
	\begin{center}
	\includegraphics[width=0.7\textwidth]{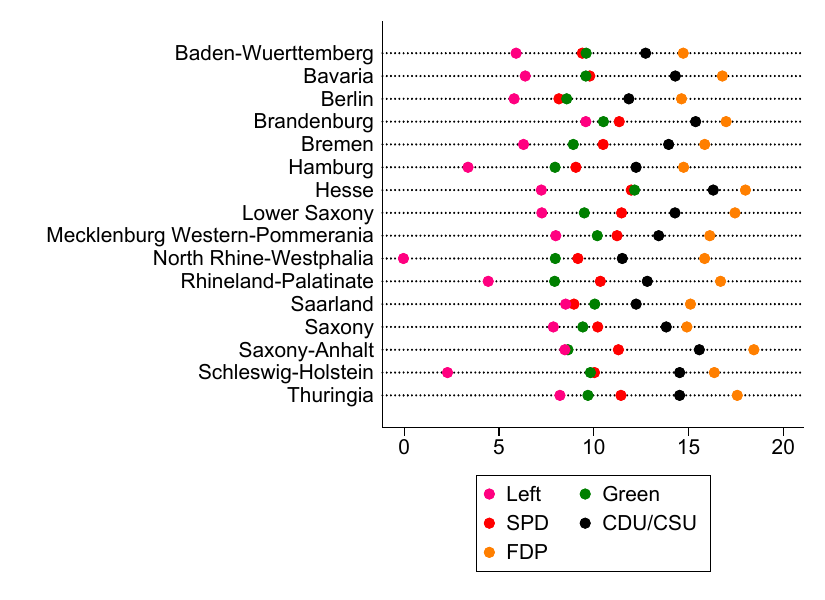}
	\end{center}
	\footnotesize{\textit{Notes:} The figure shows how social policy positions differ between the political parties in every individual state. Standardized scores measure party positions for social policies. Small values of the economic policy measure mean that the parties advocate a large size and scope of government, large values mean that the parties advocate market-oriented economic policies. The data are from Appendix B in \citet{braeuninger2020book}.}
	\label{fig: economic policy positions}
\end{figure}

Economic policy positions differed less than social policy positions between the political parties in every individual state (see the economic policy positions in Figure \ref{fig: economic policy positions}). In BW, the Greens scored 9.6, and the CDU scored 12.74, which indicates that the Greens and the CDU agreed more on economic than social policy positions. This is, however, true for many other German states as well. The difference in the economic policy positions between the Greens and the CDU in BW ($12.74 - 9.6 = 3.14$) is, for example, quite similar to the difference between the Greens and the CDU in Berlin ($11.85 - 8.57 = 3.28$). Overall, the differences between the Greens/SPD and the CDU/FDP in BW are comparable to the corresponding differences in the other German states.

We focus on the manifestos in the year 2011 or the elections before 2011 to relate to issues before the 2011 election in BW as closely as possible. The results hardly change when we investigate manifestos over the period 1990-2019; see Figures 5.1 and 5.2 in  \citet{braeuninger2020book}. 

One may also want to compare differences in party positions of leftwing and rightwing coalitions and examine benchmark cases. For example, in North Rhine-Westphalia, a leftwing SPD/Green government succeeded a rightwing CDU/FDP government in the year 2010. Comparing the average economic policy position of the previous CDU/FDP government ($(11.50 + 15.85)/2 = 13.68$) and the succeeding leftwing SPD/Green government ($(9.16 + 7.97)/2 = 8.57$) yields a difference of $13.68 - 8.57 = 5.11$. In BW in 2011, the economic policy position of the previous CDU/FDP government and the succeeding Green/SPD government were ($(12.74 + 14.72)/2 = 13.73$) and ($(8.58 + 8.16)/2 = 8.37$). The difference of $13.73 - 8.37 = 5.46$ is quite comparable to the difference in North Rhine-Westphalia.\footnote{The social policy positions of the CDU/FDP and SPD/Greens governments in North Rhine-Westphalia were ($(11.17 + 8.01)/2 = 9.59$) and ($(6.46 + 8.34)/2 = 7.40$). The difference is $9.59 - 7.40 = 2.19$. The social policy positions of the CDU/FDP and the Green/SPD governments in BW were ($(10.96 + 7.95)/2 = 9.46$) and ($(7.18 + 4.36)/2 = 5.77$). The difference is $9.46 - 5.77 = 3.69$.}

In sum, party competition and changes in government in BW are comparable with party competition in the other German states.

\section{Robustness: restricting the donor to states where nuclear power plants closed in 2011}
\label{app: nuclear power plants}

The federal government decided to abolish nuclear energy after the Fukushima nuclear disaster: 6 of the 17 nuclear power plants were closed in 2011, one of which in BW. Removing this energy capacity from the system was possible because Germany is well connected in Europe's energy market and imports/exports quite a bit of its energy \citep{grossietal2017}. 

A concern for our identification strategy is that abolishing nuclear power plants may have had different effects on states with and without nuclear power plants. If this is the case, the units in the donor pool without nuclear power plants may not be not suitable controls. Therefore, we investigate the robustness of our findings for the energy and environmental outcomes, restricting the donor pool to the four states in which nuclear power plants were closed in 2011: Bavaria, Hesse, Lower-Saxony, and Schleswig-Holstein.

\begin{figure}[H]
    \caption{Results for energy outcomes based on restricted donor pool}
    \begin{center}
        \vspace{-0.4cm}
    \begin{subfigure}[b]{0.325\textwidth}
         \centering
         \caption{CO2 emissions}
         \includegraphics[width=\textwidth]{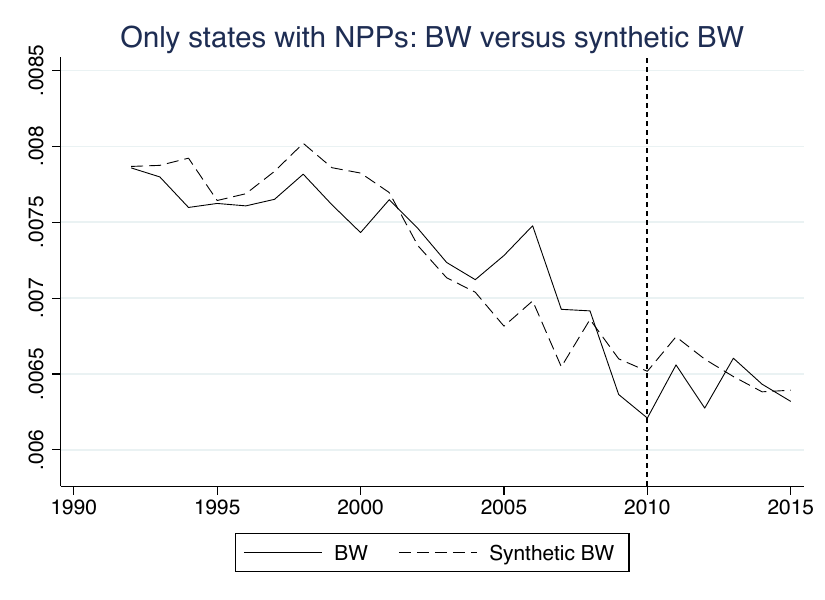}
    \end{subfigure}
    \hfill
    \begin{subfigure}[b]{0.325\textwidth}
         \centering
         \caption{Mineral oil}
         \includegraphics[width=\textwidth]{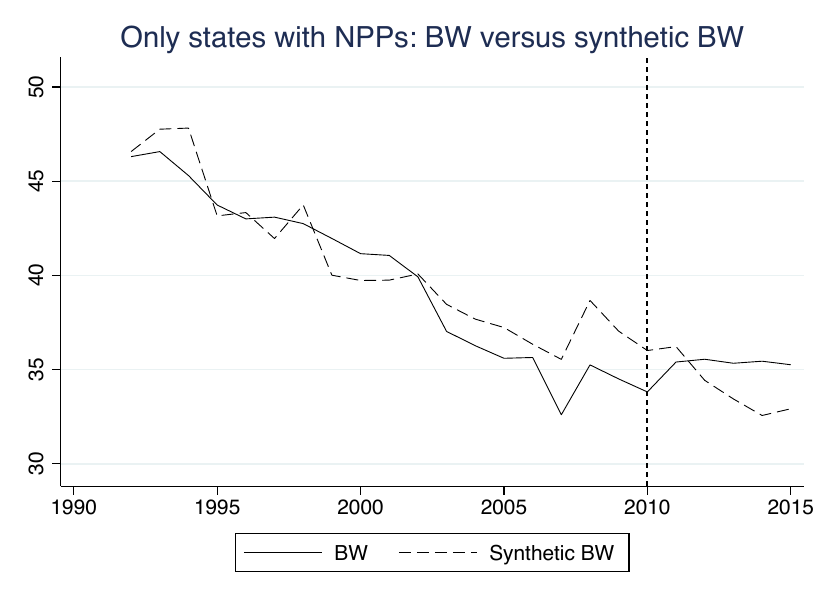}
         \end{subfigure}
    \begin{subfigure}[b]{0.325\textwidth}
        \centering
        \caption{Coal}
        \includegraphics[width=\textwidth]{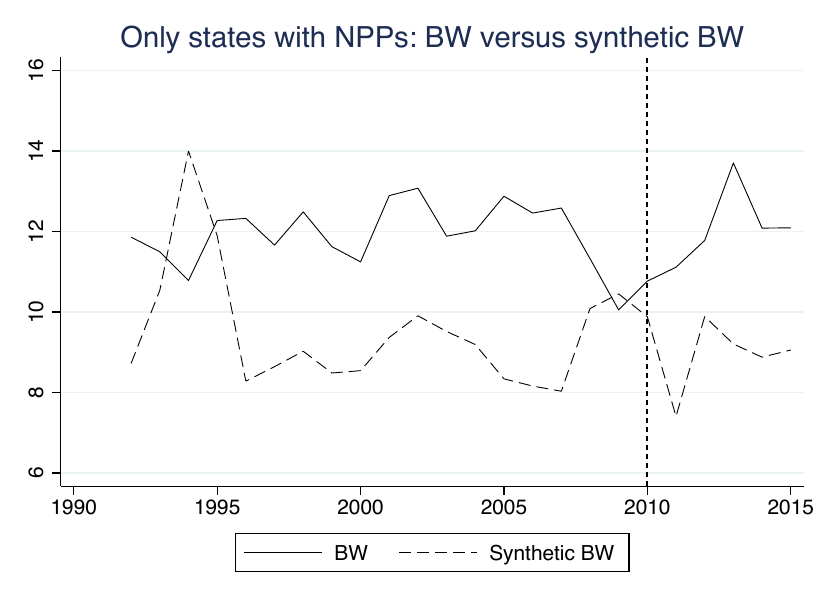}
    \end{subfigure}
    \begin{subfigure}[b]{0.325\textwidth}
        \centering
        \caption{Renewable energies}
         \includegraphics[width=\textwidth]{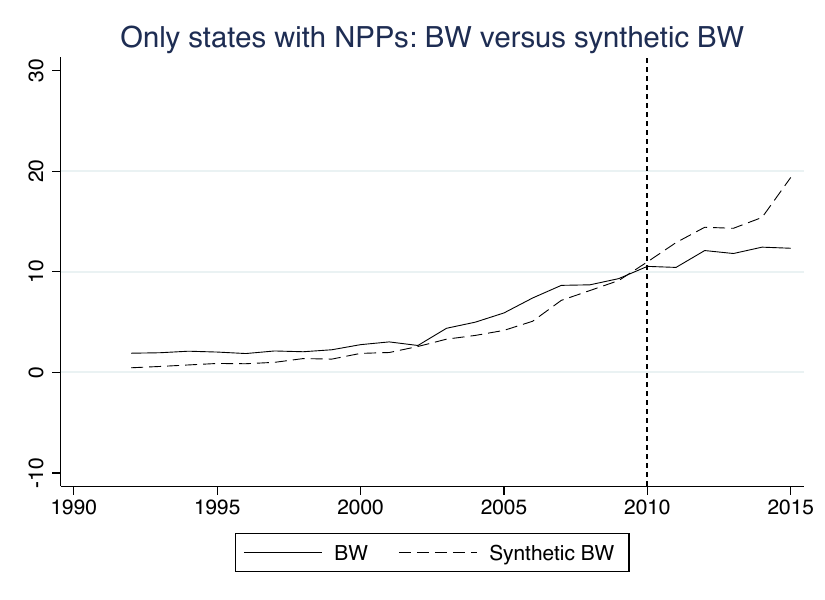}
    \end{subfigure} 
    \begin{subfigure}[b]{0.32\textwidth}
        \centering
        \caption{Wind energy}
         \includegraphics[width=\textwidth]{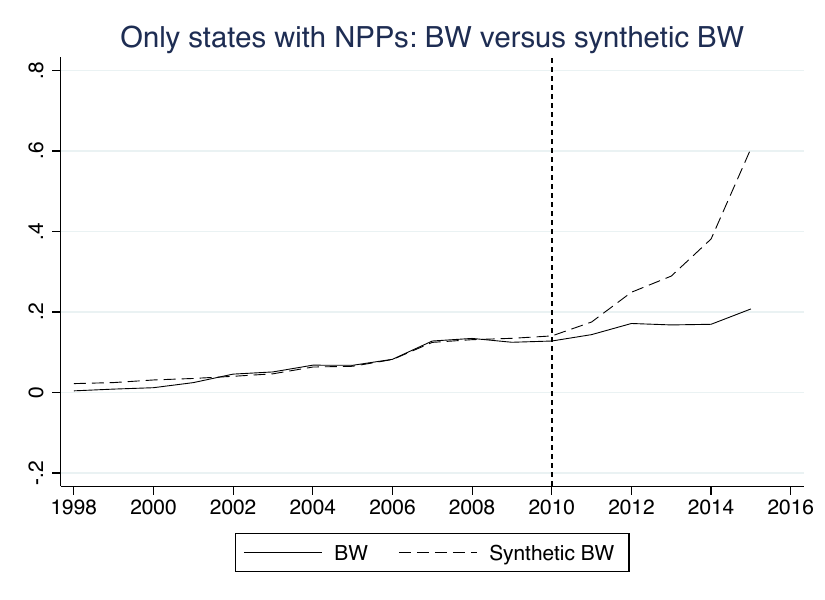}
    \end{subfigure}   
    \begin{subfigure}[b]{0.325\textwidth}
        \centering
        \caption{Solar energy}
         \includegraphics[width=\textwidth]{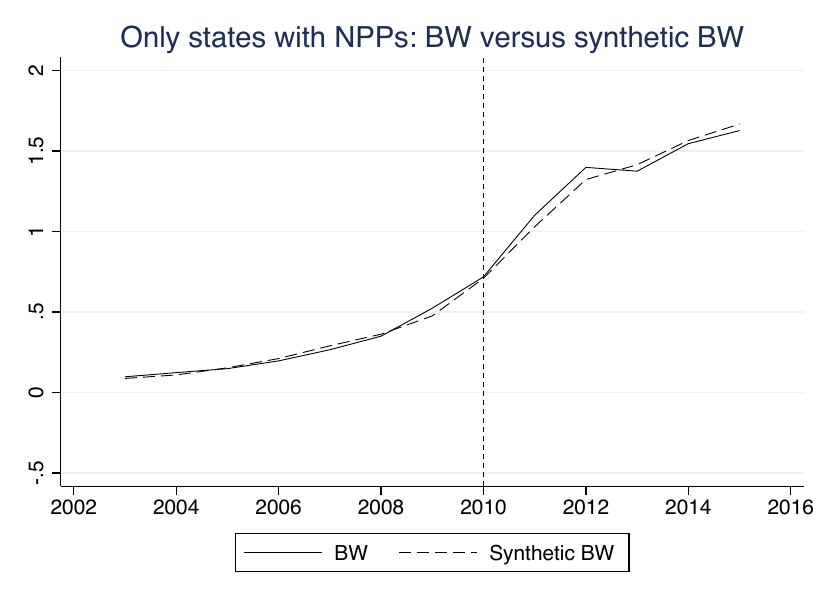}
    \end{subfigure} 
    \end{center}     
	\footnotesize{\textit{Notes:} CO2 emissions are measured in 1000t per inhabitant. The energy outcomes are measured as a share of primary energy usage. The data are from the State Working Committee for Energy Balances and the Federal Environment Agency.}
\label{fig: robustness restricted donor pool}
\end{figure}

Figure \ref{fig: robustness restricted donor pool} compares BW to the synthetic BW obtained based on the restricted donor pool. The results of the robustness check corroborate our main findings. We emphasize that despite the pre-treatment fits getting worse for some outcomes (as expected), the negative effect of the Green government on the share of wind energy remains almost unchanged. The reason is that Bavaria gets most of the weight in the synthetic BW based on both the restricted and the original donor pool.

\section{Additional results}

\subsection{Nature reserves and landscape conservation areas}
\label{app: add energy}

\begin{figure}[H]
\caption{SC results: nature reserves and landscape conservation areas}

\vspace{-0.4cm}

	\begin{center}
	 \begin{subfigure}[b]{\textwidth}
    \caption{Nature reserves}
		\includegraphics[width=0.325\textwidth,trim=0 0cm 0 0cm]{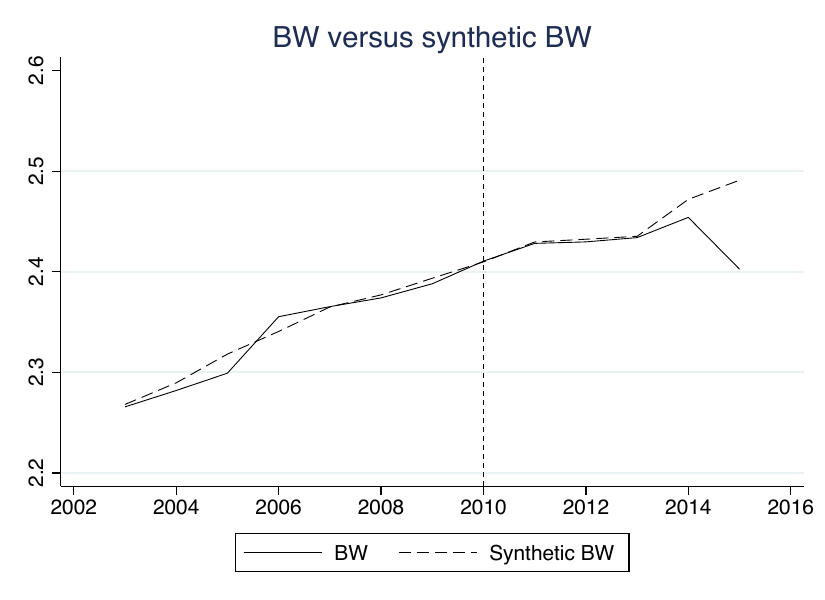}
		\includegraphics[width=0.325\textwidth,trim=0 0cm 0 0cm]{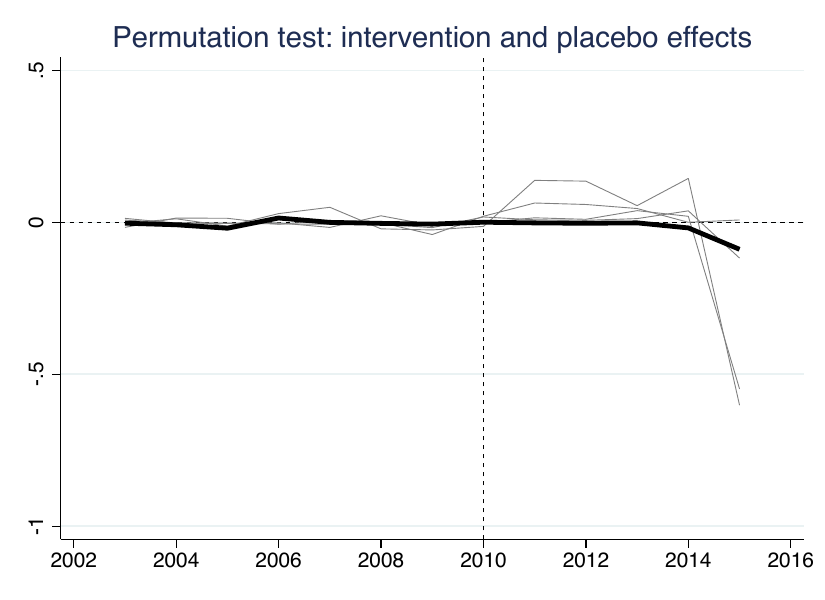}
		\includegraphics[width=0.325\textwidth,trim=0 0cm 0 0cm]{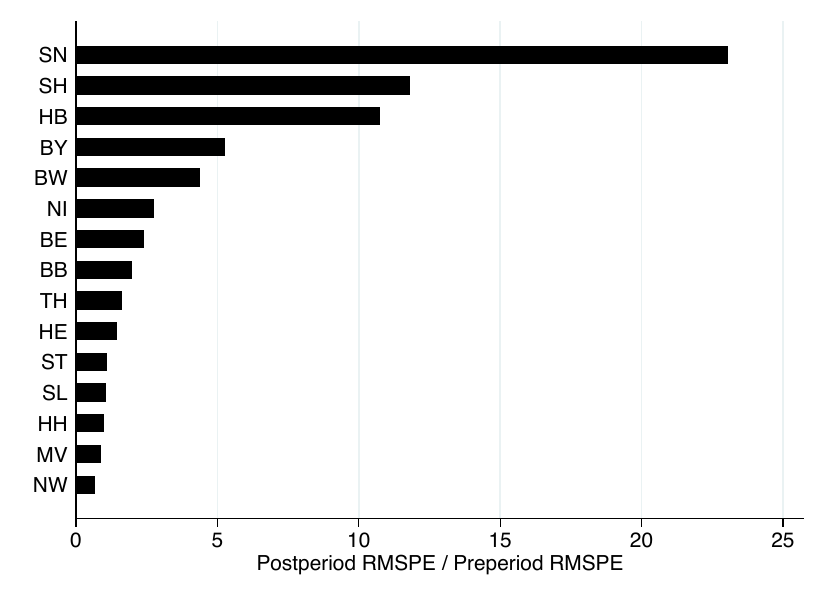}
\end{subfigure}
	 \begin{subfigure}[b]{\textwidth}
	\caption{Landscape conservation areas}
		\includegraphics[width=0.325\textwidth,trim=0 0cm 0 0cm]{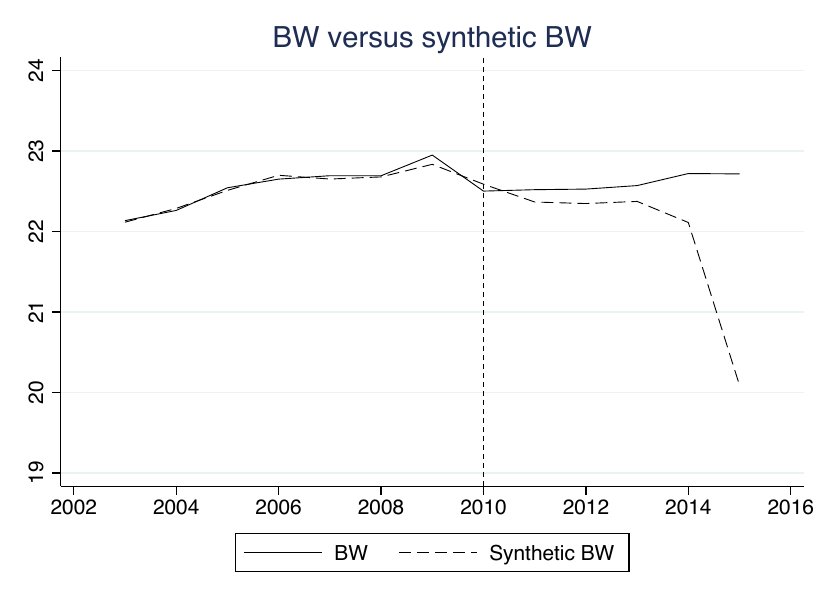}
		\includegraphics[width=0.325\textwidth,trim=0 0cm 0 0cm]{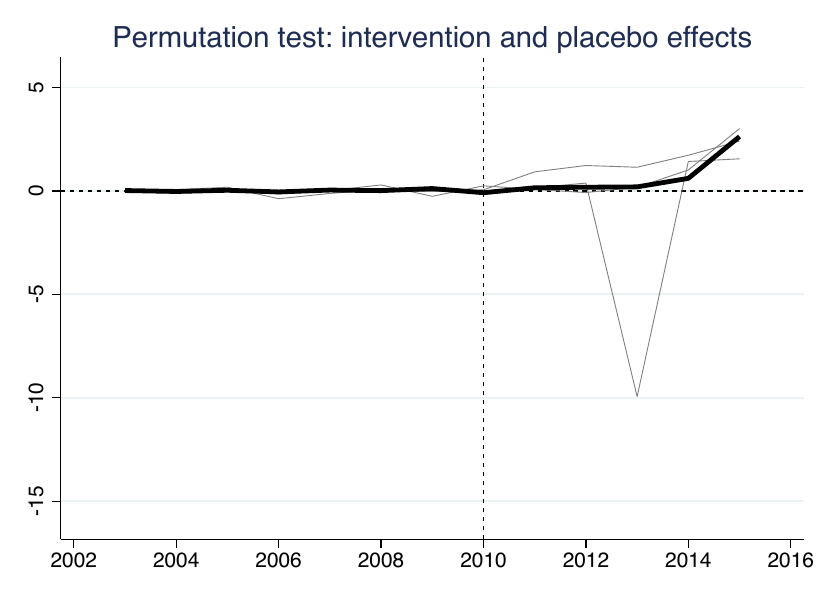}
		\includegraphics[width=0.325\textwidth,trim=0 0cm 0 0cm]{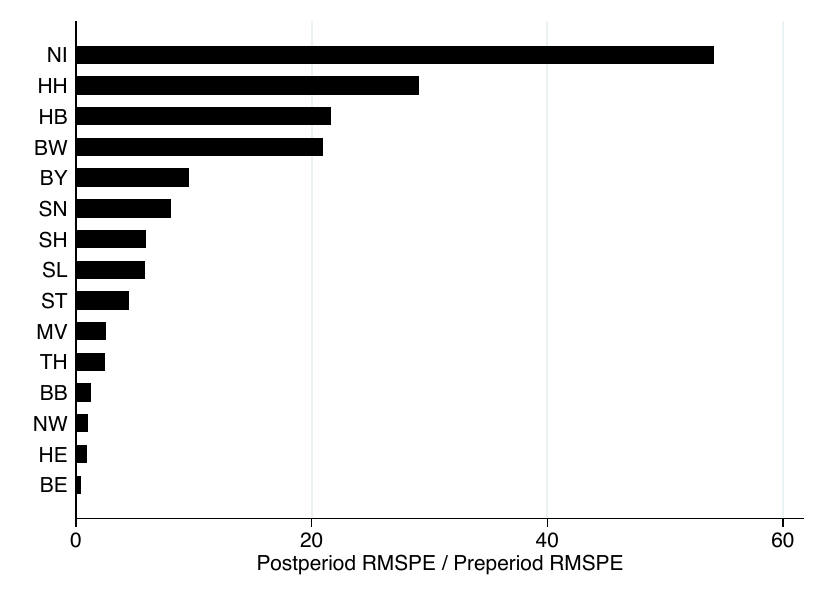}
\end{subfigure}

	\end{center}

	\footnotesize{\textit{Notes:} The center panel excludes states for which the pre-treatment MSPE is at least 10 times larger than BW's pre-treatment MSPE. The data are from the Federal Agency for Nature Conservation.}
	\label{fig: Nature reserve}
\end{figure}

\subsection{Secondary and lower secondary schools}
\label{app: add education}

\begin{figure}[H]
	\caption{SC results: secondary and lower secondary schools}
	\vspace{-0.4cm}
	
	\begin{center}
	\begin{subfigure}[b]{\textwidth}
	\caption{Number of students in secondary schools}
		\includegraphics[width=0.325\textwidth,trim=0 0cm 0 0cm]{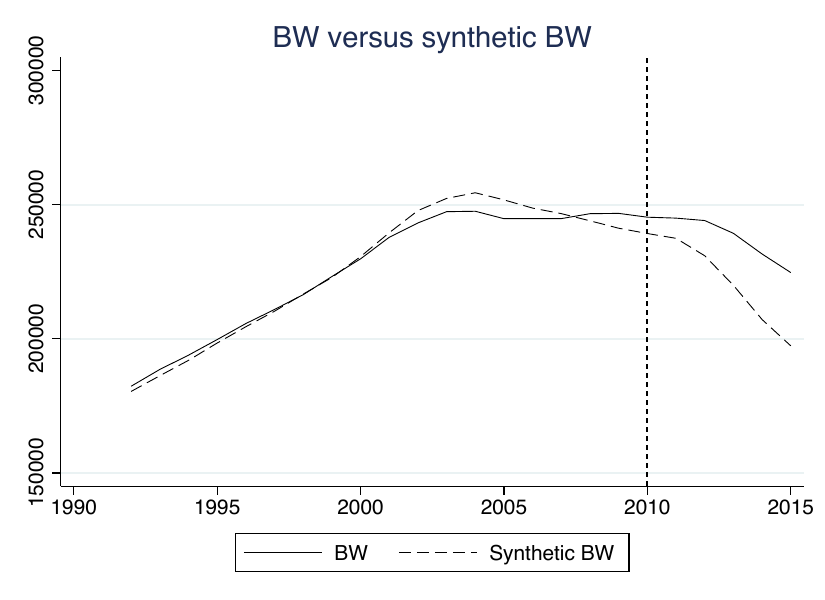}
        \includegraphics[width=0.325\textwidth,trim=0 0cm 0 0cm]{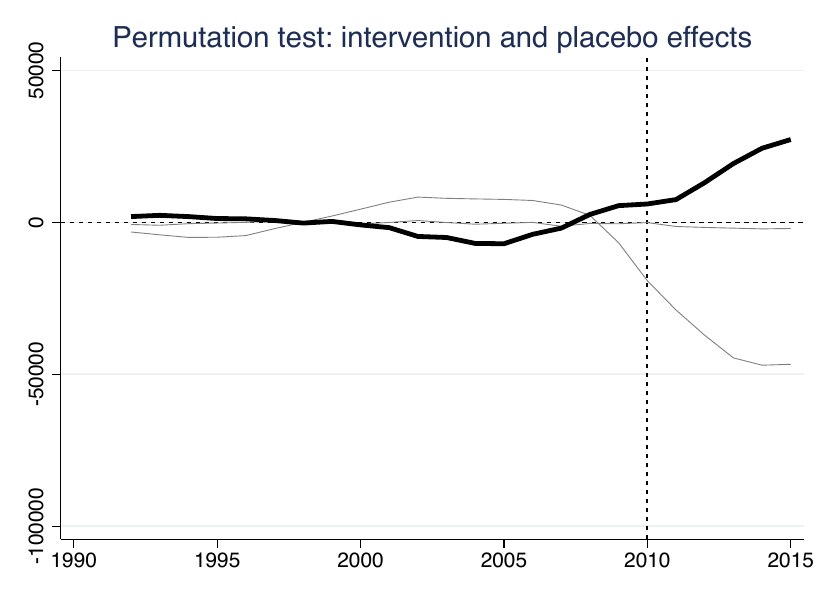}
        \includegraphics[width=0.325\textwidth,trim=0 0cm 0 0cm]{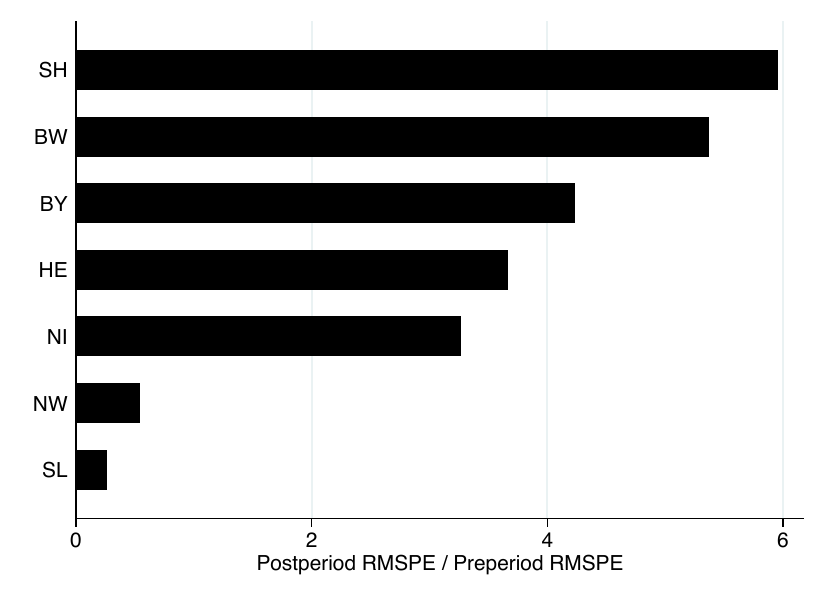}
	\end{subfigure}
	\begin{subfigure}[b]{\textwidth}
		\caption{Number of students in lower secondary schools}

		\includegraphics[width=0.325\textwidth,trim=0 0cm 0 0cm]{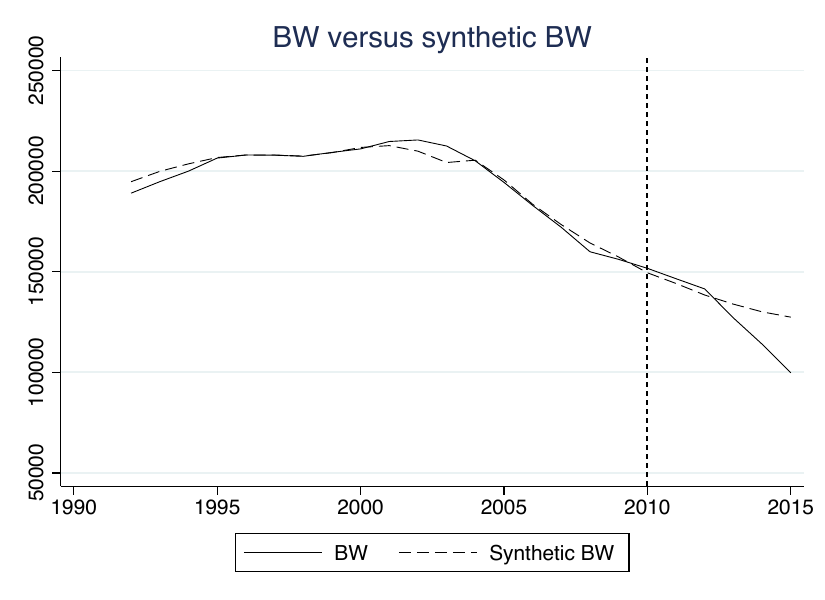}
        \includegraphics[width=0.325\textwidth,trim=0 0cm 0 0cm]{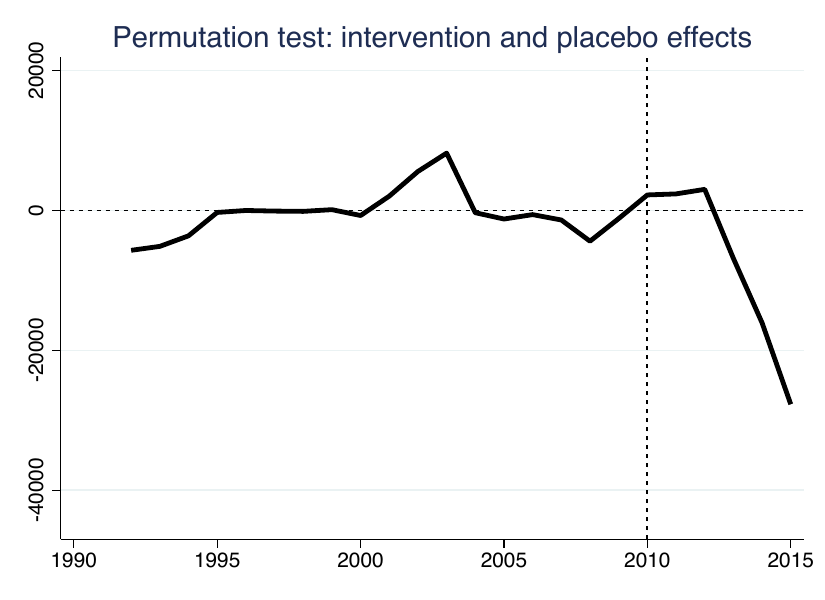}
        \includegraphics[width=0.325\textwidth,trim=0 0cm 0 0cm]{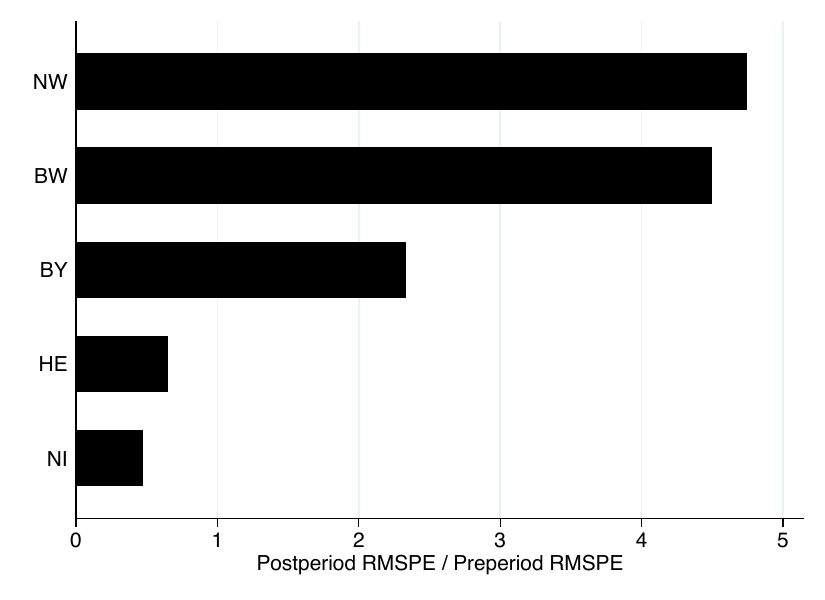}
	\end{subfigure}
		\end{center}
	 \footnotesize{\textit{Notes:} The center panel excludes states for which the pre-treatment MSPE is at least 10 times larger than BW's pre-treatment MSPE. The data are from the Standing Conference of the Ministers of Education and Cultural Affairs of the Laender.}
	\label{fig: Secondary schools: students}
\end{figure}

\newpage

\section{Synthetic control weights}
\label{app: sc weights}

\begin{table}[H]
\tiny
\centering
\caption{SC weights: landscape protection outcomes}

\begin{tabular}{lcc}
\\
\toprule
          & \multicolumn{1}{c}{\textbf{Nature reserves}} & \multicolumn{1}{c}{\textbf{Landscape conservation areas}}
          \\ \cmidrule{2-3}     \\
    Brandenburg              & 0      & 0              \\
    Berlin                   & 0      & 0.191          \\
    Bavaria                  & 0.836  & 0              \\
    Bremen                   & 0      & 0              \\
    Hesse                    & 0.025  & 0              \\
    Hamburg                  & 0      & 0              \\
    Mecklenburg-Pommerania   & 0.004  & 0              \\
    Lower-Saxony             & 0.040  & 0              \\
    North Rhine-Westphalia   & 0      & 0.011          \\
    Schleswig-Holstein       & 0      & 0.402          \\
    Saarland                 & 0      & 0              \\
    Saxony                   & 0      & 0              \\
    Saxony-Anhalt            & 0.065  & 0.382          \\
    Thuringia                & 0.031  & 0.014          \\
    \\
\bottomrule
\end{tabular}
\label{tab: tablelandscapeweights}
\end{table}

\begin{table}[H]
\centering
\tiny\caption{SC weights: macroeconomic outcomes}
\begin{tabular}{lcc}

\toprule
          & \multicolumn{1}{c}{\textbf{GDP per employee}} & \multicolumn{1}{c}{\textbf{Unemployment rate}}
          \\ \cmidrule{2-3}     \\
Brandenburg				&	0				&	0	\\
Berlin					&	0				&	0	\\
Bavaria					&	0.321			&	1	\\
Bremen					&	0				&	0	\\
Hesse					&	0.356			&	0	\\
Hamburg					&	0				&	0	\\
Mecklenburg-Pommerania	&	0				&	0	\\
Lower-Saxony			&	0.176		    &	0	\\
North Rhine-Westphalia	&	0.051	    	&	0	\\
Schlewig-Holstein		&	0				&	0	\\
Saarland				&	0.096	    	&	0	\\
Saxony					&	0				&	0	\\
Saxony-Anhalt			&	0		    	&	0	\\
Thuringia				&	0				&	0	\\
\\
\bottomrule
\end{tabular}
\label{tab: tablemacroweights}
\end{table}

\begin{table}[H]
\tiny
\centering
\caption{SC weights: education outcomes}

\begin{tabular}{lcccc}
\\
\toprule
          & \multicolumn{1}{c}{\textbf{Comprehensive schools}} 
          & \multicolumn{1}{c}{\textbf{High schools}} 
          & \multicolumn{1}{c}{\textbf{Secondary schools}}   &\multicolumn{1}{c}{\textbf{Lower secondary schools}}
          \\ \cmidrule{2-5}     \\
    Brandenburg              & 0      & 0.081 &       &        \\
    Berlin                   & 0.019  & 0     &       &        \\
    Bavaria                  & 0.451  & 0.836 & 0.188 & 0.557  \\
    Bremen                   & 0.036  & 0     &       &        \\
    Hesse                    & 0.011  & 0     & 0.268 & 0.163  \\
    Hamburg                  & 0      & 0     &       &        \\
    Mecklenburg-Pommerania   & 0      & 0     &       &        \\
    Lower-Saxony             & 0      & 0.083 & 0.026 & 0.279  \\
    North Rhine-Westphalia   & 0      & 0     & 0.519 & 0      \\
    Schleswig-Holstein       & 0      & 0     & 0     &        \\
    Saarland                 & 0      & 0     & 0     &        \\
    Saxony                   &        & 0     &       &        \\
    Saxony-Anhalt            & 0.255  & 0     &       &        \\
    Thuringia                & 0.228  & 0     &       &        \\
    \\
\bottomrule
\end{tabular}
\label{tab: tableeducationweights}
\end{table}

\section{Ministers cabinet Kretschmann}
\label{app:cabinet}

\begin{table}[H]
\caption{Ministers cabinet Kretschmann}
\begin{footnotesize}

\begin{center}
\begin{tabular}{  p{5cm}  p{5cm}   p{1.5cm}  } 
 \toprule
 
 \textbf{Name} & \textbf{Ministry} & \textbf{Party} \\
 \hline
 
 Winfried Kretschmann & Prime Minister & Greens \\ 
 \hline
Nils Schmid	& Deputy Prime Minister; Finance and Economics	&  SPD\\
\hline
Silke Krebs&State Ministry& Greens \\
\hline
Reinhold Gall&Interior& SPD\\
\hline
Rainer Stickelberger&Justice& SPD\\
\hline
Gabriele Warminski-Leitheusser (until 01/07/2013) Andreas Stoch (from 01/23/2013) & Education, Youth and Sports & SPD\\
\hline
Theresia Bauer&	Science, Research and Culture & Greens\\
\hline
Katrin Altpeter&	Labor, Social Affairs, Families, Women and Senior Citizens&SPD\\
\hline
Winfried Hermann&	Transport and Infrastructure& 	Greens\\
\hline
Franz Untersteller&	Environment, Climate Protection and Energy&	 Greens\\
\hline
Alexander Bonde&	Rural Affairs and Consumer Protection&	 Greens\\
\hline
Bilkay \"Oney&	Integration&	SPD\\
\hline
Peter Friedrich&	Bundesrat, Europe and International Affairs	&SPD\\

 \bottomrule
\end{tabular}
\end{center}
\footnotesize{\textit{Notes:} The table shows the ministers of Kretschmann's cabinet. The SPD had more ministers than the Greens because Kretschmann was prime minister. However, two Green secretaries of state were also members of the government and were entitled to vote in the government. The Greens therefore had a majority in the government (eight Greens against seven SPD politicians). They also had one more seat than the SPD in the state parliament.}
\label{tab: cabinet}
\end{footnotesize}
\end{table}

\section{Dependent variables and data sources}

\begin{table}[H]
\scriptsize
\begin{center}
\caption{Dependent variables and data sources}

\begin{tabular}{lll}
\\
\toprule
          & \multicolumn{1}{c}{\textbf{Data period}} & \multicolumn{1}{c}{\textbf{Source}}
          \\ \cmidrule{2-3}     \\
    CO2 emissions in 1000t per inhabitant & 1992-2015 & State Working Committee for Energy Balances \\
    Mineral oil (\% of primary energy usage) & 1992-2015 & State Working Committee for Energy Balances \\
    Coal (\% of primary energy usage)& 1992-2015 & State Working Committee for Energy Balances \\
    Renewable energies & 1992-2015 & State Working Committee for Energy Balances \\
    (\% of primary energy usage) \\
    Wind energy (\% of primary energy usage) & 1998-2015 & State Working Committee for Energy Balances \\
    Solar energy (\% of primary energy usage) & 2003-2015 & State Working Committee for Energy Balances \\
    Nature reserves & 2003-2015 & Federal Agency for Nature Conservation \\
    (area in \% of state's overall area) \\
    Landscape conservation area & 2003-2015 & Federal Agency for Nature Conservation \\
    (area in \% of state's overall area) \\
        GDP per employee & 1992-2015 & Regional Accounts VGRdL \\
    Unemployment rate & 1992-2015 & Federal Agency of Work \\
    Comprehensive schools: students & 1992-2015 & CMC \\
    High schools: students & 1992-2015 & CMC \\
    Secondary schools: students & 1992-2015 & CMC \\
    Lower secondary schools: students & 1992-2015 & CMC \\
\\  
\bottomrule
\end{tabular}
\label{tab: depvars}

\end{center}

\end{table}

\newpage

\end{appendix}

\end{document}